\documentclass[reprint,prd,letterpaper,superscriptaddress,amsmath,amssymb,aps] {revtex4-1}

\usepackage{graphicx}
\usepackage{dcolumn}
\usepackage{todonotes} 
\usepackage{amsmath} 
\usepackage{bm}
\usepackage{siunitx}
\usepackage{subfigure}
\usepackage{pifont}
\usepackage{xcolor}
\newcommand{\cmark}{\ding{51}}

\usepackage[implicit=true, debug=false, 
plainpages=false, colorlinks=true, 
linkcolor=black, urlcolor=black, citecolor=black, filecolor=black, 
hyperindex=true]{hyperref}

\newcommand{\atUC}{\affiliation{Dept. of Physics, Enrico Fermi Inst., Kavli Inst. for Cosmological Physics, Univ. of Chicago, Chicago, IL 60637.}}
\newcommand{\atUCLA}{\affiliation{Dept. of Physics and Astronomy, Univ. of California, Los Angeles, Los Angeles, CA 90095.}}
\newcommand{\atOSU}{\affiliation{Dept. of Physics, Center for Cosmology and AstroParticle Physics, Ohio State Univ., Columbus, OH 43210.}}
\newcommand{\atUH}{\affiliation{Dept. of Physics and Astronomy, Univ. of Hawaii, Manoa, HI 96822.}}
\newcommand{\atNTU}{\affiliation{Dept. of Physics, Grad. Inst. of Astrophys.,\& Leung Center for Cosmology and Particle Astrophysics, National Taiwan University, Taipei, Taiwan.}}
\newcommand{\atKU}{\affiliation{Dept. of Physics and Astronomy, Univ. of Kansas, Lawrence, KS 66045.}}
\newcommand{\atWU}{\affiliation{Dept. of Physics, McDonnell Center for the Space Sciences, Washington Univ. in St. Louis, MO 63130.}}
\newcommand{\atSLAC}{\affiliation{SLAC National Accelerator Laboratory, Menlo Park, CA, 94025.}}
\newcommand{\atUD}{\affiliation{Dept. of Physics, Univ. of Delaware, Newark, DE 19716.}}
\newcommand{\atUCL}{\affiliation{Dept. of Physics and Astronomy, University College London, London, United Kingdom.}}
\newcommand{\atJPL}{\affiliation{Jet Propulsion Laboratory, Pasadena, CA 91109.}}

\newcommand{\atCalPoly}{\affiliation{Physics Dept., California Polytechnic State Univ., San Luis Obispo, CA 93407.}}
\newcommand{\atMPL}{\affiliation{Max-Planck-Institut f\"{u}r Kernphysik, 69117 Heidelberg, Germany.}}
\newcommand{\atBRUS}{\affiliation{Astrophysical Institute, Vrije Universiteit Brussel, Pleinlaan 2, 1050, Brussels, Belgium.}}
\newcommand{\atLSST}{\affiliation{LSST, 950 North Cherry Avenue, Tucson, AZ 85721.}}
\newcommand{\atUCSD}{\affiliation{Center for Astrophysics and Space Sciences, Univ. of California, San Diego, La Jolla, CA 92093.}}
\newcommand{\atMoscow}{\affiliation{Moscow Engineering Physics Institute, Moscow, Russia.}}

\newcommand{\combinedBackgroundEstimates}{$0.7^{+0.5}_{-0.3}$}

\begin{document} 
\bibliographystyle{apsrev4-1} 
\title{Constraints on the diffuse high-energy neutrino flux from the third flight of ANITA}

\author{P.~W.~Gorham}\atUH 
\author{P.~Allison}\atOSU
\author{O.~Banerjee}\atOSU
\author{L.~Batten}\atUCL 
\author{J.~J.~Beatty}\atOSU 
\author{K.~Bechtol}\atUC\atLSST
\author{K.~Belov}\atJPL 
\author{D.~Z.~Besson}\atKU\atMoscow
\author{W.~R.~Binns}\atWU 
\author{V.~Bugaev}\atWU 
\author{P.~Cao}\atUD 
\author{C.~C.~Chen}\atNTU 
\author{C.~H.~Chen}\atNTU
\author{P.~Chen}\atNTU 
\author{J.~M.~Clem}\atUD 
\author{A.~Connolly}\atOSU 
\author{L.~Cremonesi}\atUCL 
\author{B.~Dailey}\atOSU 
\author{C.~Deaconu}\atUC 
\author{P.~F.~Dowkontt}\atWU 
\author{B.~D.~Fox}\atUH 
\author{J.~W.~H.~Gordon}\atOSU
\author{C.~Hast}\atSLAC
\author{B.~Hill}\atUH 
\author{S.~Y.~Hsu}\atNTU
\author{J.~J.~Huang}\atNTU
\author{K.~Hughes}\atUC\atOSU
\author{R.~Hupe}\atOSU 
\author{M.~H.~Israel}\atWU 
\author{K.~M.~Liewer}\atJPL
\author{T.~C.~Liu}\atNTU 
\author{A.~B.~Ludwig}\atUC 
\author{L.~Macchiarulo}\atUH 
\author{S.~Matsuno}\atUH 
\author{C.~Miki}\atUH 
\author{K.~Mulrey}\atUD\atBRUS
\author{J.~Nam}\atNTU
\author{C.~Naudet}\atJPL
\author{R.~J.~Nichol}\atUCL
\author{A.~Novikov}\atKU\atMoscow
\author{E.~Oberla}\atUC 
\author{S.~Prohira}\atKU
\author{B.~F.~Rauch}\atWU
\author{J.~M.~Roberts}\atUH\atUCSD
\author{A.~Romero-Wolf}\atJPL
\author{B.~Rotter}\atUH
\author{J.~W.~Russell}\atUH 
\author{D.~Saltzberg}\atUCLA
\author{D.~Seckel}\atUD
\author{H.~Schoorlemmer}\atUH\atMPL
\author{J.~Shiao}\atNTU
\author{S.~Stafford}\atOSU
\author{J.~Stockham}\atKU
\author{M.~Stockham}\atKU
\author{B.~Strutt}\atUCLA 
\author{M.~S.~Sutherland}\atOSU
\author{G.~S.~Varner}\atUH
\author{A.~G.~Vieregg}\atUC
\author{S.~H.~Wang}\atNTU
\author{S.~A.~Wissel}\atCalPoly

\collaboration{ANITA Collaboration}\noaffiliation

\date{\today}

\begin{abstract}
  The Antarctic Impulsive Transient Antenna (ANITA), a NASA long-duration balloon payload, searches for radio emission from interactions of
  ultra-high-energy neutrinos in polar ice. The third flight of ANITA (ANITA-III)
  was launched in December 2014 and completed a 22-day flight. We present the
  results of three
  analyses searching for Askaryan radio emission of
  neutrino origin.  In the most sensitive of the analyses, we find one event in the signal region on
  an expected background of \combinedBackgroundEstimates. Though consistent with the background estimate, the event remains compatible with a neutrino hypothesis even after additional post-unblinding scrutiny. 

\end{abstract}

\maketitle

\section{Introduction}

Ultra-high-energy ($>$ 100 PeV) neutrinos are expected to be produced from interactions of high-energy cosmic rays with cosmic microwave background photons~\cite{BZ}.
The low expected flux~\cite{kotera} and small cross section
require monitoring
an immense volume of a dense material
for successful detection.  Coherent Cherenkov emission in the radio regime ({\it i.e.} 
Askaryan emission~\cite{askaryan}) from neutrino-induced
showers in radio-transparent dense dielectric media such as ice provides a viable mechanism
for achieving a large enough detector volume for detection of the highest
energy neutrinos. The expected signal is broadband up to
a cutoff frequency of $\sim$~GHz and the emitted power scales quadratically
with shower energy. 

The Antarctic Impulsive Transient Antenna (ANITA), a NASA long-duration balloon 
payload~\cite{instrument}, is an array of high-gain antennas that monitors the
Antarctic ice sheet for impulsive, broadband neutrino and cosmic-ray-induced radio emission.  
ANITA is not only sensitive to Askaryan emission from neutrino-induced showers
in ice, but can also observe geomagnetic emission from
extensive air showers (EAS) induced by cosmic rays or decaying $\tau$ leptons created by $\tau$ neutrino interactions~\cite{anita1CR,mysteryEvent}.
The analyses described here are optimized to look for neutrino-induced Askaryan 
emission, but are also sensitive to the EAS channel.  The EAS channel is a useful sideband region for these analyses, which is a region of phase space adjacent to the neutrino signal region, and useful for determining cut values and estimating efficiencies and backgrounds. Due to the direction of Earth's magnetic field in Antarctica, EAS emission is mainly horizontally polarized. Askaryan emission visible to ANITA is mostly vertically polarized for Standard Model cross sections, due to preferential Fresnel effects as the radio pulse propagates through the ice surface. 

\section{Experimental technique}
 \begin{figure*} 
   \includegraphics[width=0.9\textwidth]{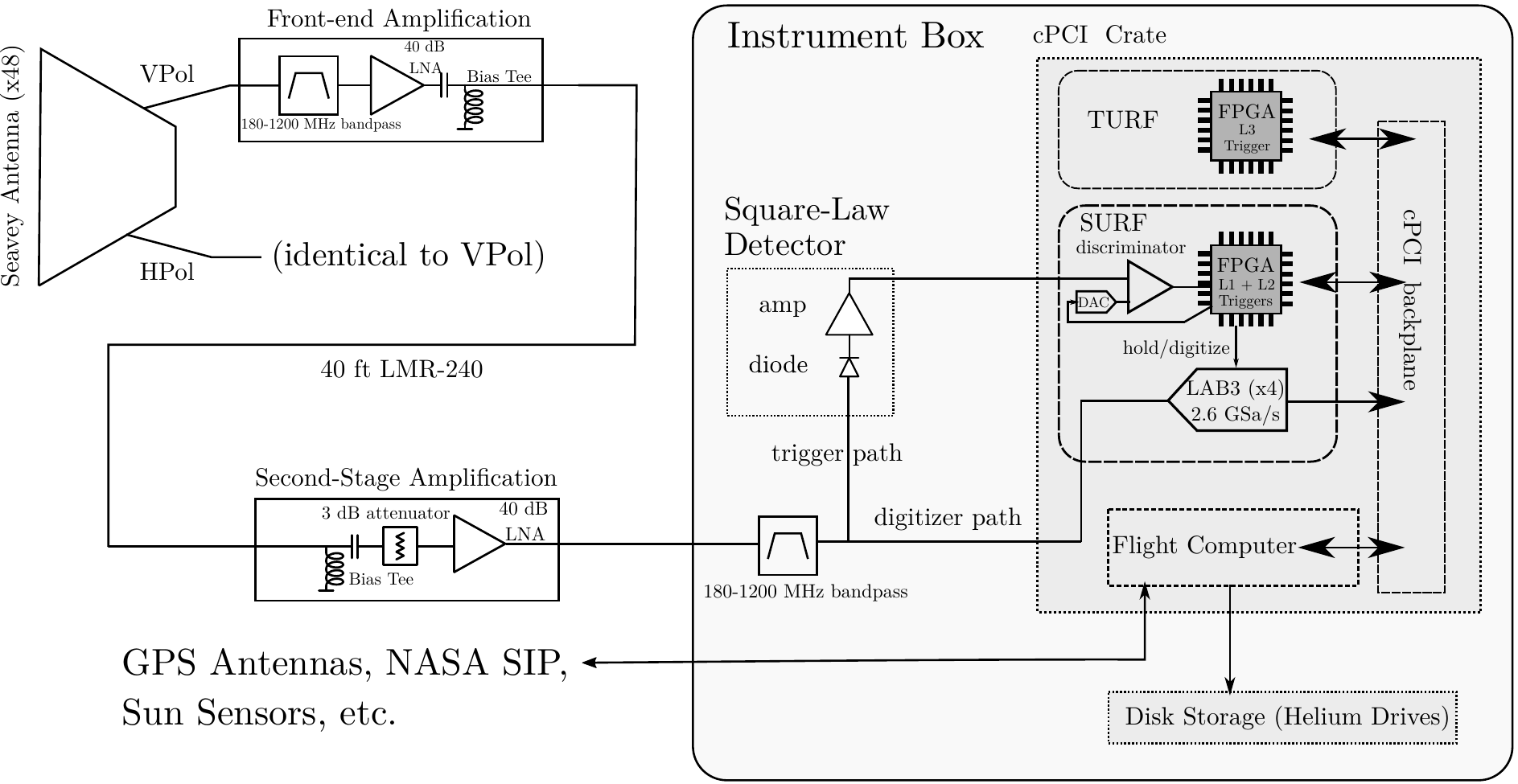} 

   \caption{A schematic diagram of the ANITA-III instrument.  Signals from
     48~dual-polarized, quad-ridge horn antennas are fed into a bandpass
     filter, through a low-noise amplifier, and then through a second-stage
     amplifier. Then, the signals are each split into two parts.     The
     trigger path signal passes through a tunnel diode square-law detector and
     amplifier before being compared to a threshold that forms a first-stage
     (channel-level) trigger.  If a global trigger (a coincidence between multiple channels) is issued, the signal from the
     digitizer path is digitized using a LAB3 switched capacitor array
     sampling at 2.6~GSa/sec, recorded on the flight computer, and stored to
     disk.  The first levels of the trigger and digitization are performed by a
     custom board called a Sampling Unit for Radio Frequencies (SURF).  The Trigger Unit for Radio Frequencies (TURF) collects lower-level trigger information from the entire payload
     to form global triggers. The NASA Support Instrument Package (SIP) is responsible for telemetry. }
     \label{fig:anita3}
   \end{figure*}
  
The third flight of the ANITA experiment, ANITA-III, launched 
on 18 December 2014.  The instrument is similar to 
the previous two ANITA payloads~\cite{anita1,anita2}.  The primary upgrades from ANITA-II are 
the addition of 8 more antennas and a low-frequency antenna (ALFA) aimed at enhancing detection of EAS signals, the implementation of a new impulsive full-band-only trigger in both horizontal and vertical polarizations, and the use
of new lower-noise radio-frequency amplifiers. Here we briefly describe the instrument, flight,
and calibration procedures. 

\subsection{The ANITA-III instrument}

    \begin{figure}
    \includegraphics[width=\columnwidth]{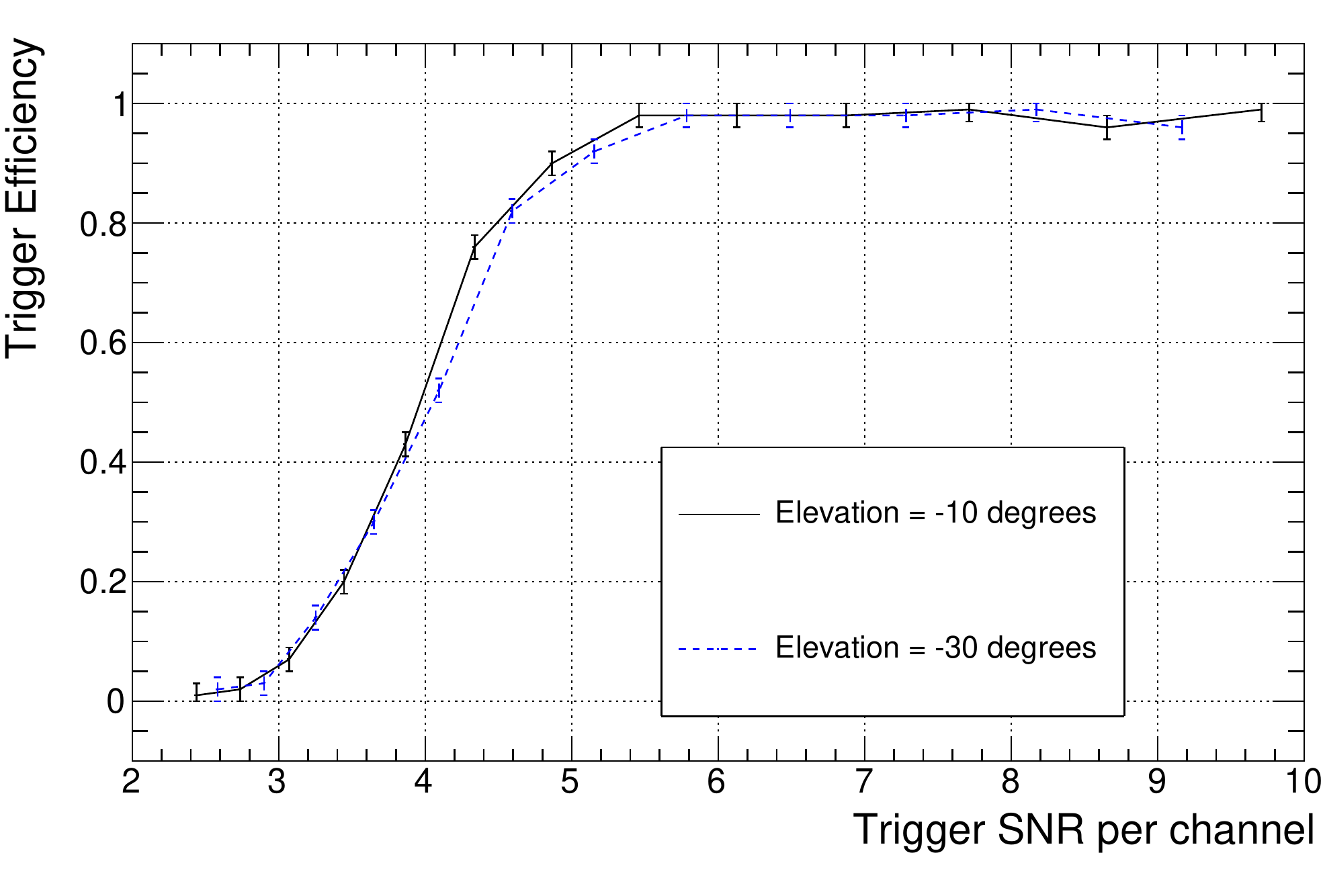} 
    \caption{Trigger efficiency vs. voltage signal-to-noise ratio (SNR), derived from lab measurements of injected signals in three adjacent azimuthal sectors. Efficiencies are shown for sets of delays between antennas corresponding to two different elevation angles. } 
     \label{fig:trigger_snr} 
    \end{figure}

  A schematic of the ANITA-III instrument and data acquisition system is depicted in
  Fig.~\ref{fig:anita3}. Forty-eight dual-polarization quad-ridge horn antennas from Antenna Research Associates, Inc. are arranged in a three-ring vertical cylindrical pattern to form 96 wideband 
  (180~MHz-1200~MHz) channels. Each ring has 16 antennas, and each grouping of three
  antennas (top, middle, bottom) are azimuthally aligned, forming 16 azimuthal
  sectors.   The signal from each channel is bandpass-filtered and then amplified by a 
  custom-built low-noise amplifier, which is 
  adjacent to the antenna, and then split into trigger and
  digitization paths after a second stage of amplification. Antenna temperatures are typically $\sim$ 130~K and the noise temperatures for the front-end filters and amplifiers are $\sim$100~K. 

  The trigger path uses a custom tunnel diode as a fast square-law
  detector. The tunnel diode output is compared to a dynamically-adjusted
  threshold to determine if a channel-level (first-level) trigger should be issued. Unlike
  previous ANITA payloads, first-level triggers are based solely on the total power within approximately 10-ns
  coincident windows
  in each channel, not the frequency content of the signal~\cite{instrument}.  The
  trigger thresholds are adjusted in real time 
  to keep the first-level
  trigger rate approximately at its target rate, which for ANITA-III was 450 kHz.

A second-level trigger condition is imposed at the level of each azimuthal sector and is satisfied by a coincidence of
  two or more channels in a single polarization within the sector.  If a first-level trigger is issued for a given channel, a 
  coincidence window opens during which another channel in the
  same azimuthal sector satisfying the first-level condition would
  generate a second-level trigger. The size of the coincidence window
  depends on the rings involved in the trigger, set by the expectation for up-going signals: 16~ns for the bottom ring, 
  12~ns for the middle ring, 
  	and 4~ns for the top ring.

  The third-level (global) trigger is generated by the coincidence of second-level
  triggers occurring in the same polarization in adjacent azimuthal sectors.  A global
  trigger will cause the digitized signals to be read out, 
  assuming the four-deep digitizer buffer is not full. Over the course of the flight, the average deadtime incurred from full digitizer buffers was 13\%.  Because the triggers for horizontal and vertical polarizations
  operate independently, it is possible to have a simultaneous trigger for both.  The global
  trigger rate over the course of the flight for ANITA-III was approximately 50~Hz. The trigger efficiency as a function of voltage signal-to-noise ratio (SNR) in the trigger chain, derived from lab measurements, is shown in Fig.~\ref{fig:trigger_snr}. The trigger efficiency reaches 50\% at a voltage SNR of $4.0\sigma$.

  The digitizer path uses LAB3~\cite{lab4} switched capacitor array digitizers
  with a mean sample rate of 2.6~GSa/s. Each channel has four 260-sample analog
  buffers to minimize deadtime. 

  In addition to the science triggers generated by the trigger logic described
  above, there are triggers generated either by the payload computer or a pulse
  per second signal from the onboard GPS devices. These provide a set of
  minimum-bias triggers to help assess the noise environment during flight.

  To prevent a portion of the payload from triggering too often and
  monopolizing all available digitizer buffers, a trigger mask is automatically
  enabled by the flight computer if an azimuthal sector's global trigger
  rate exceeds a configurable threshold.  This allows ANITA to dynamically 
  mask channels from the trigger that are subject at any given time 
  to significant anthropogenic (man-made) noise from locations in Antarctica. Because of satellite
  interference in ANITA-III,
  throughout most 
  of the flight the channels that are North-facing at a given time are masked. 

  The ANITA payload rotates freely. Two independent ADU5 differential GPS
  units are used to measure the payload attitude and position.   
  Power is supplied and controlled with photo-voltaic panels, a bank of batteries, 
  and a charge controller.  
  Telemetry is available during the flight through Iridium, TDRSS (when
  available), and a line-of-sight system when near McMurdo Station, the largest base of operations in Antarctica. 

  \subsection{The ANITA-III flight}
  \begin{figure} 
    \includegraphics[width=\linewidth]{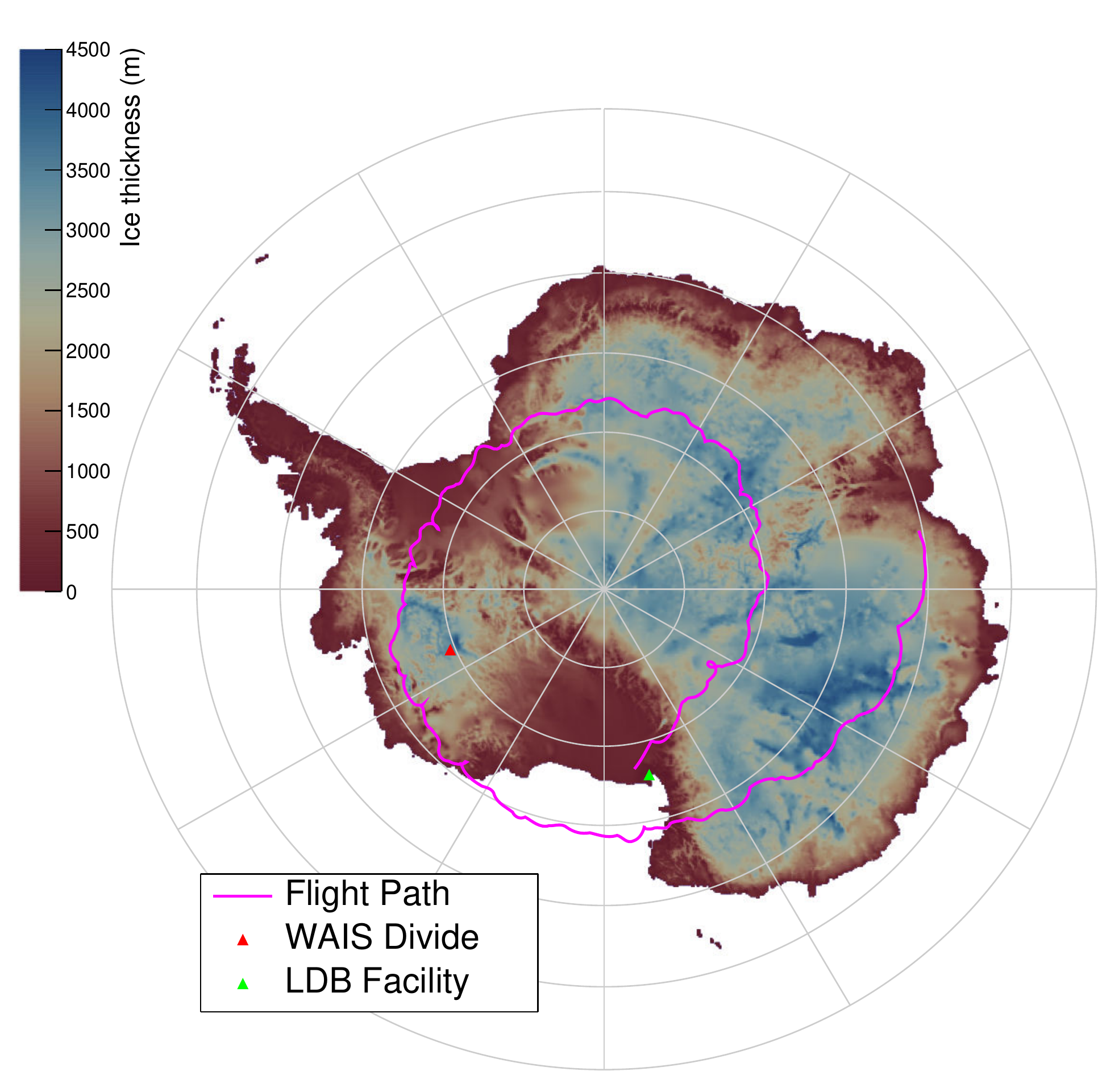} 
    \caption{The ANITA-III flight path is shown on top of a map of ice depth\cite{bedmap2}.
      The location of a high-voltage radio impulse generator used as a calibration source (WAIS Divide) and the launch site (LDB Facility) are also shown.
    }\label{fig:flightpath}
  \end{figure}

  ANITA-III launched from the NASA long-duration balloon facility on the Ross Ice
  Shelf near McMurdo station on December $18^{\textrm{th}}$, 2014.  ANITA-III flew for
  22 days before termination on January $9^{\textrm{th}}$, 2015.  The flight path is shown in Fig.~\ref{fig:flightpath}.  
  The hard disks and flight hardware were recovered with the 
  aid of the Australian Antarctic Division from nearby Davis station. 

 High-voltage impulsive calibration signals were 
 sent to ANITA-III from the
  launch site and from an autonomous high-voltage calibration pulser deployed at
  WAIS divide. These field pulsers are referenced to GPS time
  to facilitate identification. The data from the WAIS pulser proved particularly useful since
  ANITA-III passed close enough to be triggered over 100,000 times by the
  pulser. 

  ANITA requires extensive calibration of each digitizer in order to make full
  use of the precision timing information. In addition, a temperature
  correction must be applied to account for changes in clock frequency as a function
  of temperature. A
  detailed summary of these calibration procedures is provided in~\cite{BStruttThesis,BRotterThesis}.  

\section{Analysis Methods}

Of the over seventy million science triggers captured during the ANITA-III
flight, at most a few events of neutrino origin are expected.  The
threshold-riding trigger on the instrument is set so that the vast majority of
those events are thermal noise, the level of which in turn dictates ANITA's
threshold.  The majority of the remaining events are anthropogenic transient
and continuous-wave (CW) emission and occasional impulsive emission from the
on-board electronics, which we call {\it payload blasts}.
 
After reviewing the backgrounds to the search and the simulation tools, we will
briefly summarize the three neutrino searches performed.  Additional
detail for each analysis is provided in appendices. 

\subsection{Classes of backgrounds}

The vast majority of recorded ANITA-III events are random fluctuations of thermal noise due to ANITA's threshold-riding trigger. 
The typical antenna temperature for ANITA-III is $\sim130$~K, from a
combination of the sky and the ice that is in their field of view. 

Anthropogenic CW from terrestrial 
transmitters or satellites will also trigger ANITA-III.  
In particular, the 260 and 380~MHz communication bands
used by various satellites are a dominant
cause of science triggers for ANITA-III.
Even events that triggered on an impulsive neutrino-like signal can have 
a significant contribution from CW sources, which complicates analysis.

Self-triggered payload blasts are impulsive radio-frequency emission 
generated by electronics on the ANITA payload. 
Although ANITA electronics are heavily shielded to prevent leakage of electromagnetic interference (EMI) from the payload, some unknown source of self-interference still appears sporadically in the data. Payload blasts are characterized by non-planar wavefront geometry (since they originate 
from very close to the antennas), a distinct frequency spectrum, and are typically much
stronger in the bottom and middle rings of antennas than the top ring (also due to their
being local to the payload). 

Isolated, broadband impulsive anthropogenic emission from the ground and thermal noise fluctuations  
that by chance reconstruct as coming from the continent are both sources of background that remain
after analysis cuts are developed.  In all cases, the contribution to the
expected background in the signal region is estimated before unblinding the
search.

\subsection{Simulation} 

The primary ANITA simulation tool is \texttt{icemc}, described in detail
in~\cite{icemc}. The \texttt{icemc} program 
includes a full treatment of the ANITA trigger
and digitizer signal chain and uses the flight paths and recorded channel
thresholds in order to model the acceptance of ANITA.  It is a
weighted Monte Carlo (MC), where each generated neutrino carries a weight
corresponding to its survival probability and a phase-space factor. 

We generate a set of simulated neutrinos to characterize the efficiency of the
analyses.  The simulated neutrinos follow the maximum mixed-composition Kotera {\it et al.}~\cite{kotera} flux
model, hereafter referred to as ``Kotera", with Standard Model cross sections~\cite{crosssection}. To simulate the flight noise
environment, the trigger path was modeled with synthetic noise with levels and
spectra derived from the flight, and real minimum-bias trigger data were added
in the digitizer path. 
The choice of flux model has little effect on predicted neutrino observables. However, changes in neutrino interaction lengths, even within Standard Model bounds on cross section, affect what emission cones are visible, resulting in different observable angular and polarization distributions. 

\subsection{Summary of blind searches} 

Three independent blind neutrino analyses were performed, which we denote,
in order of completion, \textbf{A}, \textbf{B} and \textbf{C}. Analyses \textbf{A} and \textbf{B} are similar to each other and to previous ANITA analyses in using common criteria across the continent and searching for isolated events~\cite{anita1,anita2}.  Analysis \textbf{C} applies a new methodology in developing geographically-dependent
search criteria with the aim to maintain sensitivity even in regions
of ice with higher levels of anthropogenic noise.  Each neutrino search was done with at least one method of blinding: keeping hidden the region of parameter space where the signal resides, using a decimated data sample to set cuts, and/or ``salting'' the data by inserting simulated events.
Further details
are available in Appendices~\ref{sec:clustering},~\ref{sec:clustering2},
and~\ref{sec:binned}, respectively.

\begin{figure}
  \includegraphics[width=\columnwidth]{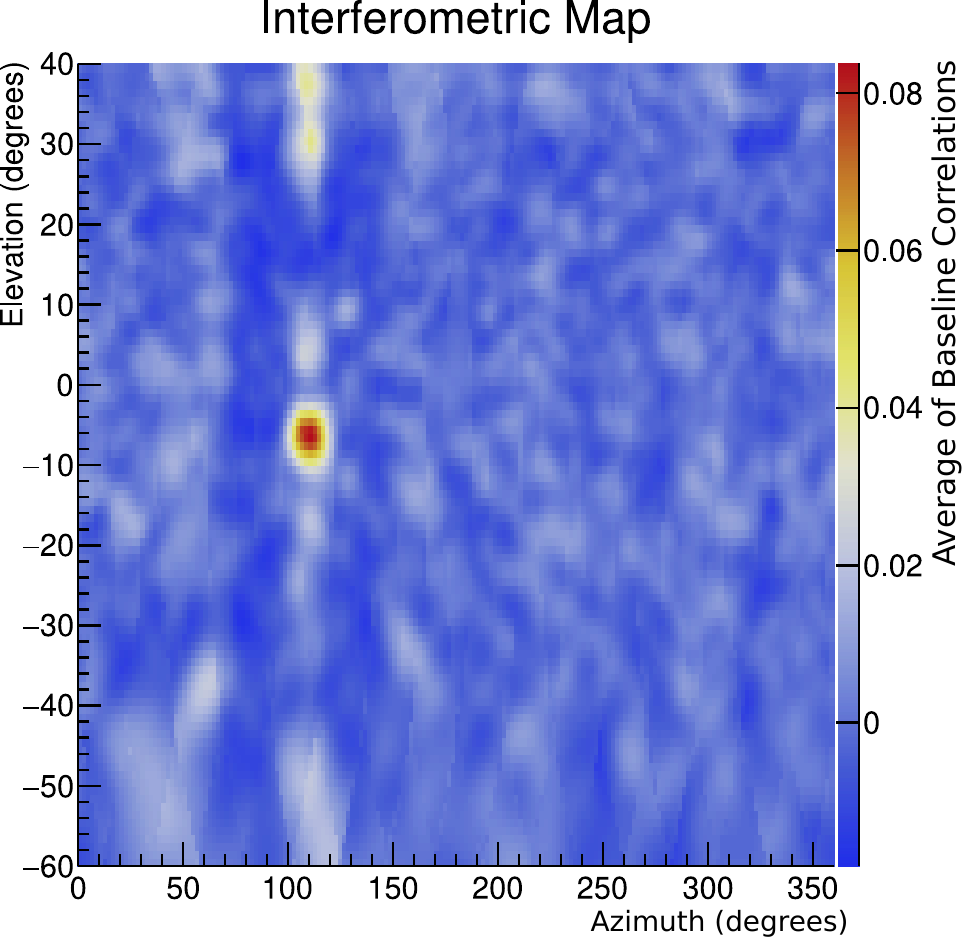}\\ 
  \vspace{0.25cm} 
  \includegraphics[width=\columnwidth]{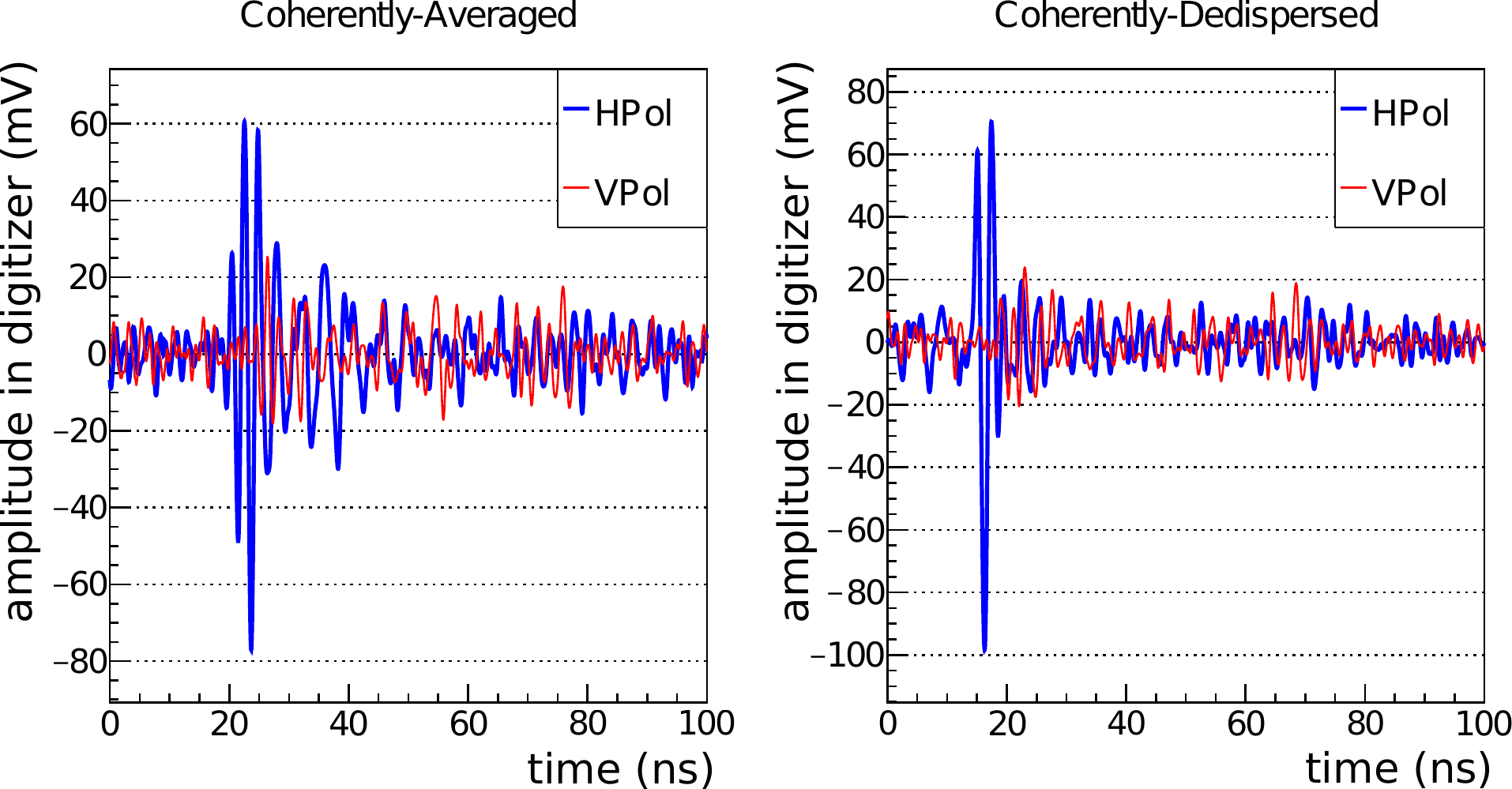} 
  \caption{An example of an interferometric map (top), a coherently-averaged waveform (bottom left), and a dedispersed coherently-averaged waveform (bottom right) for a calibration pulser event.  The color scale in the top panel corresponds to the normalized cross-correlation value.  Although we expect Askaryan neutrino signals to be mostly vertically polarized, the calibration pulser is horizontally polarized.} 
  \label{fig:interferometric_map} 
\end{figure}

Each analysis begins by filtering waveforms to mitigate undesired CW
contamination that would otherwise interfere with the analysis.  Analyses \textbf{A} and
\textbf{B} use an adaptive time-domain phasor removal technique while
\textbf{C} uses a method that 
removes CW phasors in the frequency domain~\cite{brianDaileyThesis}.

The filtered waveforms from antennas with at least a partial common field of view are correlated against each other to produce an
interferometric map~\cite{interferometric}, which indicates the apparent amount of correlated power as
a function of incoming direction. Peaks of the map are considered hypotheses of
coherent sources, for which a coherently-averaged waveform is produced. The
group delay of the instrument response can be 
removed from each waveform prior
to coherent averaging, to form a dedispersed
coherently-averaged waveform.
Fig.~\ref{fig:interferometric_map} shows an example map, a coherently-averaged
waveform, and a dedispersed coherently-averaged waveform for a calibration pulser event.

From the raw waveforms, interferometric map, and coherent waveforms, each search
computes a number of observables for each event that may be used to reject
backgrounds.  Examples of observables include the peak correlation value of the
interferometric map, the peak of a coherent waveform's analytic envelope,
measures of coherent and dedispersed waveform impulsivity,  and polarimetric
quantities.  Each search has a set of ``quality cuts'' used to remove
digitizer glitches, payload blasts, and other poor-quality
events prior to attempting to separate thermal and anthropogenic backgrounds. 

Analyses \textbf{A} and \textbf{B} use similar approaches to reject thermal and
anthropogenic noise. A multivariate linear discriminant on various observables
(different between the analyses, but much of the power in both is from measures
of impulsivity) is used to discriminate signal-like events from background. This discriminant is trained with simulated events as a signal
sample and events reconstructing above the horizontal as a non-impulsive
sideband.  Events passing the signal selection are then spatially clustered in
order to identify isolated signal-like events. Analysis~\textbf{A} projects a bivariate Gaussian distribution corresponding to the pointing resolution for each passing
event onto a map of Antarctica, creating a localization distribution on the continent, and considers the
overlap of each event's localization with the sum of the localizations of all other
events, using no \textit{a priori} information about human activity.  Analysis~\textbf{B} considers how close each event is to known locations of human activity (bases) or to the
nearest other event that passes signal-like cuts, where a fit along the continent's surface is used to find
the best mutual location for each event pair. Both searches treat horizontally and vertically-polarized
events in the same way, but only passing vertically-polarized events are in the Askaryan neutrino signal region. Passing
horizontally-polarized events contain a sample of EAS. 

Analysis \textbf{C} is complementary  in that it uses geographically-dependent selection criteria to identify events that stand out from
other events in the local noise environment. The power of this technique is in its ability to retain additional portions of the continent in the neutrino search in the presence of anthropogenic noise.

The search discretizes the continent, utilizing the HEALPix package~\cite{Gorski:2004by}, from the
start, with each  bin (about 400~km on a side) treated as an independent analysis. Cuts are optimized for the best expected limit after combining results
across bins, reflecting bin-dependent
neutrino sensitivities, noise environments, and systematic uncertainties on the background estimates.

Analysis \textbf{C} uses a 10\%
subset of the data to model the total background environment and assess the
associated systematic uncertainties on 
the background estimates.  
Based on a common appearance of
the background distributions 
across bins, we
assert that the backgrounds follow an exponential behavior in a final
cut variable. 
If an exponential fit in a bin gives a p-value below 0.05,
or insufficient data is available,
the bin is rejected from the
analysis.  In addition to the systematic uncertainties that come from the fits,
the optimization also accounts for 
a systematic uncertainty due to 
spillover of events between bins~\cite{jacobGordonThesis}.

Analysis \textbf{C} utilizes
cross-correlation values 
derived from
both linearly and circularly polarized waveforms to reject thermal noise and 
events influenced by satellite interference~\cite{samStaffordThesis}.
Treating horizontal and vertical polarizations as
separate search channels,
Analysis \textbf{C} imposes a
cut on a linear
combination of the strength of the coherent waveform and the peak 
cross correlation that is bin-dependent to distinguish thermal events from signal-like events.

Analyses \textbf{A} and \textbf{B} also set final thermal and clustering cuts by
optimizing sensitivity. Analysis~\textbf{A} estimates backgrounds with sidebands as in the on-off
problem~\cite{Rolke}, avoiding the need to assert a model for the background distributions. Analysis~\textbf{B} uses an on-off treatment for the anthropogenic
background, but an empirical model for the non-signal-like background. In both cases, 
events that reconstruct above the horizontal are used to estimate the leakage from the multivariate
discriminant.  To estimate the anthropogenic background, Analysis~\textbf{A} uses a
sideband that is sub-threshold in the multivariate discriminant while Analysis~\textbf{B} uses a
sideband of signal-like events near known bases. Analysis \textbf{A} has a
total estimated background per polarization of $0.8^{+0.6}_{-0.4}$~and
Analysis \textbf{B} expects \combinedBackgroundEstimates~per polarization. 
Analysis \textbf{C} estimates backgrounds and uncertainties
bin-by-bin that are 
about 0.1 event per bin
with $\sim$10\% systematic uncertainties.
  
The overall analysis efficiency, estimated using simulation, is 72$\pm$5\% for
Analysis \textbf{A}, 84$\pm$3\% for Analysis~\textbf{B} and $7^{+6}_{-3}$\% for Analysis
\textbf{C}, while Analysis~\textbf{C} has
more than twice its mean efficiency in some bins. 
Statistically, Analysis~\textbf{B} is the most sensitive analysis.

Analyses {\bf A} and {\bf B} choose different clustering techniques to remove anthropogenic noise: Analysis {\bf A} solely relies on event self-clustering and includes a larger event sample for clustering, while Analysis {\bf B} relies on a list of known locations of human activity as well as event self-clustering.
Analysis {\bf C} aims
to complement the other two searches by
peering into noisy as well as quiet environments using 
geographically-specific cuts, and with this aim in mind, more aggressively cut on backgrounds.
Of the simulated
neutrino events found by Analysis~\textbf{C},
25\% of them
would have been rejected by
the other two analyses.  

\section{Results}\label{sec:results}

 \begin{figure}
   \includegraphics[width=\columnwidth]{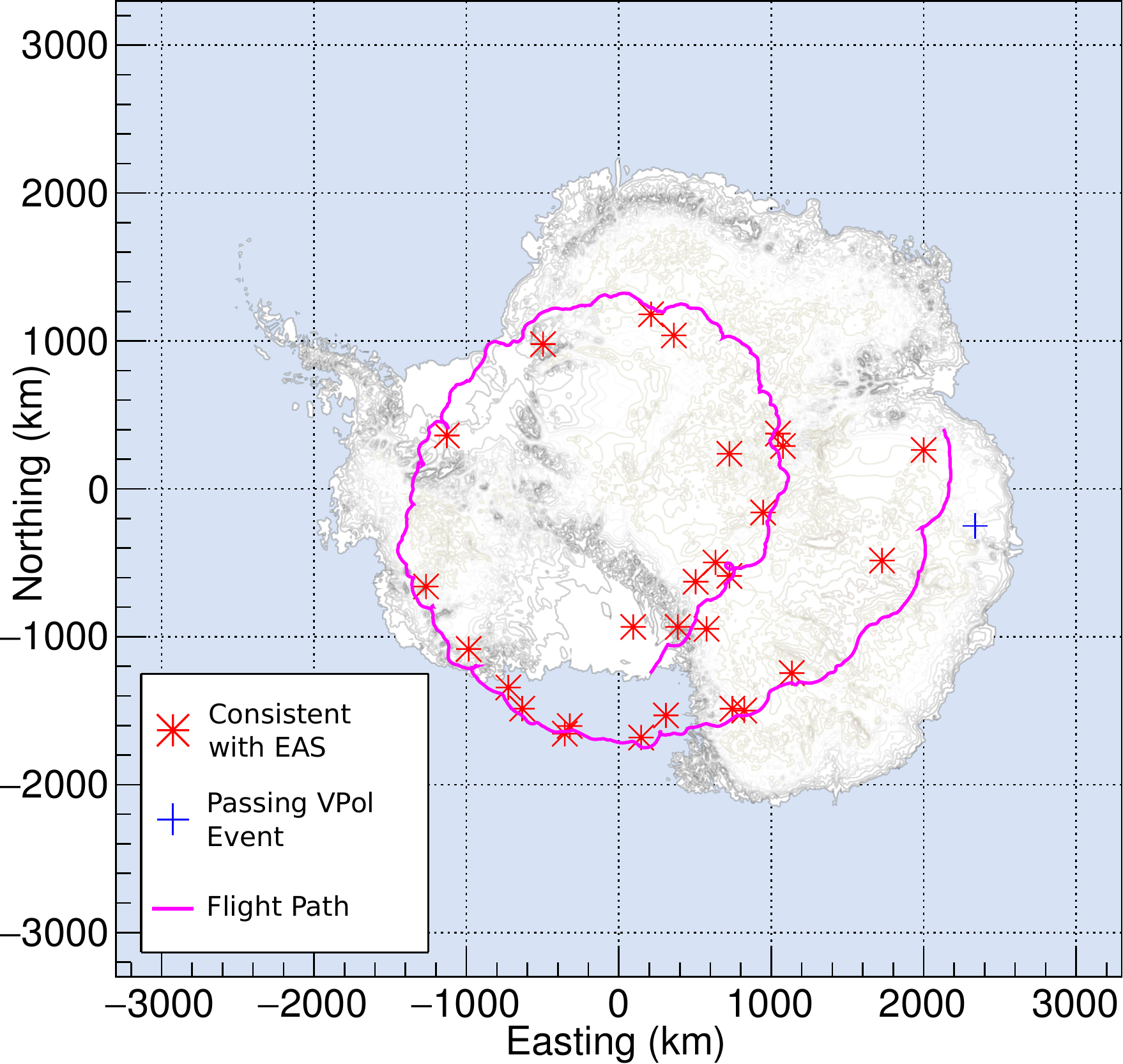} 
   \caption{Events consistent with EASs from Analyses \textbf{A}, \textbf{B}, and \textbf{C} and the event in the vertically-polarized Askaryan neutrino signal region from Analysis \textbf{B} (the most efficient of the three analyses).  Only horizontally-polarized events with a good EAS template-correlation match and consistent polarization with the local geomagnetic field are shown on the map.} 
     \label{fig:map_events} 
  \end{figure}

\begin{table} 

  \begin{tabular}{|c|c|c|c|}

 \hline
 \multicolumn{4}{|c|}{Identified by Analysis \textbf{A} }\\
 \multicolumn{4}{|c|}{Background estimate: $0.8^{+0.6}_{-0.4}$ per polarization }\\
  \multicolumn{4}{|c|}{Overall efficiency: 72$\pm5$\%}\\
 \hline
   Event & \textbf{A} & \textbf{B} & \textbf{C}  \\
 \hline
  -- & -- & -- & -- \\
 \hline
 \hline
 \multicolumn{4}{|c|}{Identified by Analysis \textbf{B}}\\
 \multicolumn{4}{|c|}{Background estimate:~\combinedBackgroundEstimates ~per polarization }\\
  \multicolumn{4}{|c|}{Overall efficiency: 84\%}\\
 \hline
   Event & \textbf{A} & \textbf{B} & \textbf{C} \\
 \hline
 ~~~83139414~~~ &   ~~~~~S~~~~ & ~~~~\cmark ~~~~ &  ~~~~P~~~~ \\
 \hline
 \hline
 \multicolumn{4}{|c|}{Identified by Analysis \textbf{C}}\\
 \multicolumn{4}{|c|}{Expect $\sim0.1$ event in each of 37 bins}\\
  \multicolumn{4}{|c|}{Effs. per bin: from few \% to $18$\% }\\
 \hline
   Event & \textbf{A} & \textbf{B} & \textbf{C} \\
 \hline
 21702154 &  S  &S  &  \cmark  \\ 
 73750661 &  S  & S &  \cmark   \\ 
 \hline
\end{tabular} 

\caption{Summary of events identified by each search in the (vertically-polarized) Askaryan neutrino search region. The analysis efficiencies on MC neutrinos and background estimates per polarization for each analysis are included. 
The one vertically-polarized event remaining in the signal region in Analysis \textbf{B} was found to be sub-threshold but isolated in Analysis \textbf{A}, and was cut by a directional cut in Analysis \textbf{C}, discussed in Appendix~\ref{sec:binned}.
All analyses find a number of events in the signal region conistent with their background estimates.
A \cmark~indicates that the event was found by a search. For events not identified, ``Q" means the event was
rejected by ``quality'' pre-selection cuts (e.g. requirements on trigger
polarization, time and \textit{a priori} elevation angle cuts), an ``S" means
the event did was sub-threshold in a signal-like selection criteria, and a ``P''
indicates the event was rejected due to its position (clustering, or, for Analysis
\textbf{C}, HEALPix bin or angular proximity to regions with geosynchronous satellites).  
  }

\label{tbl:events} 

\end{table} 

Askaryan neutrino signals are
expected to be predominantly vertically polarized for Standard Model cross sections, but all searches consider both
horizontally and vertically-polarized events. Horizontally-polarized events are not in the Askaryan neutrino signal region, but they provide a useful
cross-check on the analyses. Within the horizontally-polarized sideband region are any events from EAS from cosmic
rays and from $\tau$ leptons originating from $\nu_{\tau}$ interactions in the Earth or ice. 
\subsection{Summary of events found} 

The Askaryan neutrino 
search region is exclusively
in the vertical polarization channel.
However, we also report events identified by each analysis that pass all cuts except for the angle of linear polarization, which constitute a sample of horizontally-polarized events that we use as a validation of the relative
signal efficiencies reported each analysis.
We report on EAS candidate events in a separate paper~\cite{a3tau}.

Analysis \textbf{A} finds no events in the
Askaryan signal region and 22 events in the horizontally-polarized sideband.  Of the 22, 21 are in agreement with the expected signal shape of an EAS template and have polarization consistent with the local geomagnetic field.  The remaining event is inconsistent with an EAS hypothesis (it has both poor correlation with an EAS signal shape template and has nearly equal power in horizontal and vertical polarizations, which is not allowed by the Antarctic geomagnetic
field), but is consistent with the background estimate of $0.8^{+0.6}_{-0.4}$ in this horizontally-polarized region. Eighteen of these events that are consistent with an EAS signature were identified in a separate, dedicated EAS search~\cite{BRotterThesis}.

Analysis \textbf{B} identifies one event in the Askaryan neutrino signal region (event 83139414) and 25 events in horizontally-polarized sideband region. The event in the Askaryan neutrino signal region passed clustering cuts but was sub-threshold in Analysis~\textbf{A}. This is consistent with the slightly better analysis efficiency achieved by Analysis~\textbf{B} compared to Analysis~\textbf{A}. The 25 horizontally-polarized events include 20 of the 21 events from Analysis~\textbf{A} that are consistent with an EAS signature
and five additional events, including one separately identified by the dedicated EAS search~\cite{a3tau,BRotterThesis}.  All horizontally-polarized events that pass cuts in Analysis~\textbf{B} are consistent with emission from
EAS in both signal shape and polarization. 

Analysis \textbf{C} 
identifies two vertically-polarized events in the Askaryan neutrino signal region and seven horizontally-polarized events that pass all cuts. 
Two of the horizontally-polarized events (events 33484995 and 58592863) are also
found in Analyses~\textbf{A} and \textbf{B}, and are consistent with an EAS signature in signal shape and polarization angle. 
A third (event 48837708) is also consistent with an EAS.  The remaining four horizontally-polarized events are consistent with the background estimate.
The two events in the neutrino signal region are also consistent with the
background estimate.  We note that observing an event in each of two bins out of 37 has a negligible effect on
the flux constraints, and is one advantage 
of using a binned approach.

Table~\ref{tbl:events} lists all vertically-polarized events that pass all cuts in at least one analysis.  
The locations of all horizontally-polarized events consistent with EASs and the (vertically-polarized) event in the Askaryan neutrino signal region
identified by Analysis~\textbf{B} are shown in Fig.~\ref{fig:map_events}.
The total number of horizontally-polarized events consistent with EASs observed (27) is consistent with the EAS results from ANITA-I, scaled for the relative exposures of the two flights~\cite{anita1CR}. 

\subsection{Limit on the diffuse neutrino flux and model constraints} 

\begin{figure}
  \includegraphics[width=\columnwidth]{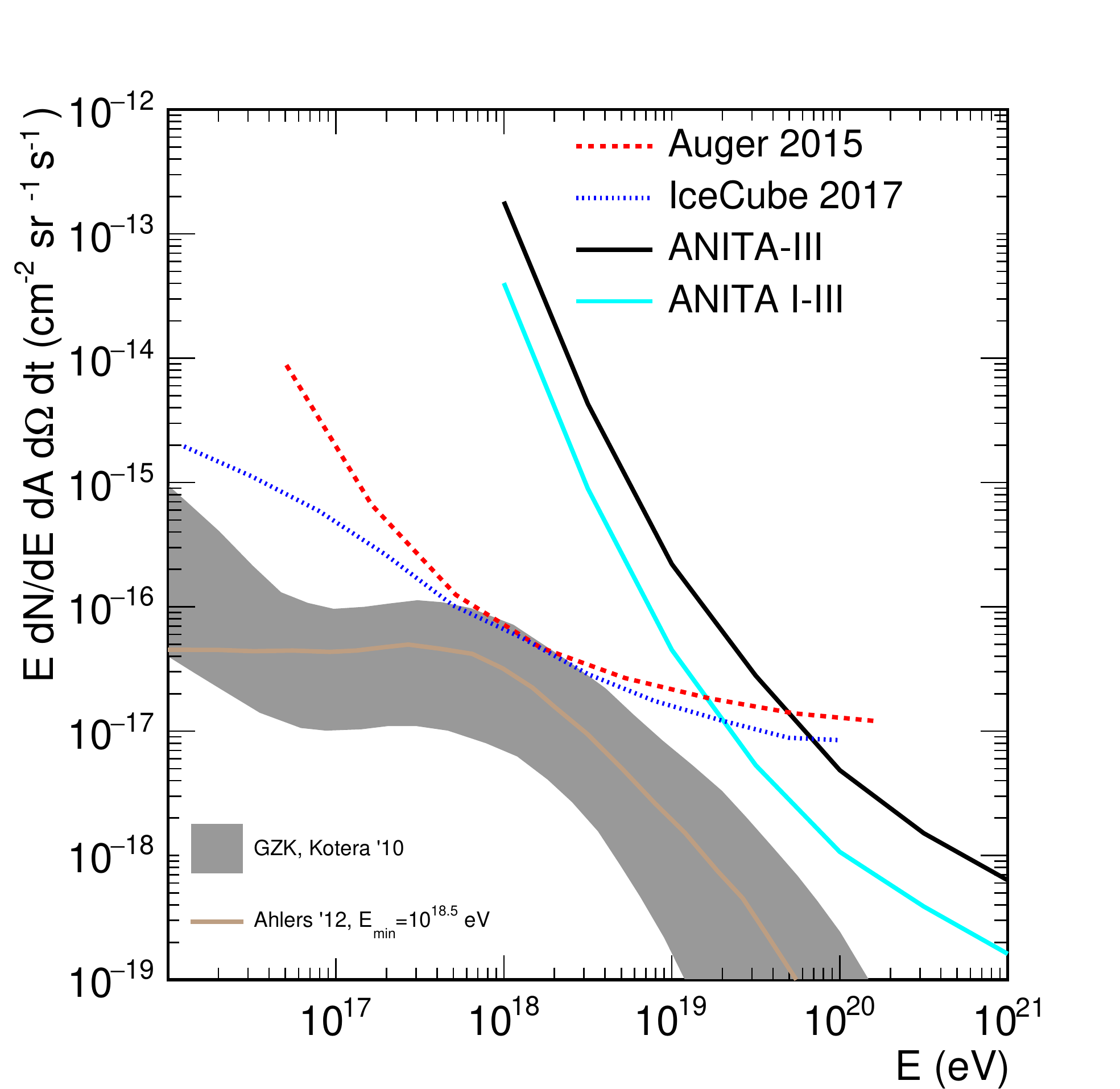} \\[0.2cm]

  \begin{tabular} 
    {l|c|c|c|c|c|c|c}
    $\log_{10}$(E(eV)) & 18 & 18.5 & 19 & 19.5 & 20 & 20.5 & 21 \\
    \hline
  A (km$^2 \cdot$sr) & 0.00038 & 0.016 & 0.31 & 2.5 & 14 & 46 & 109\\
  \end{tabular}

  \caption{ANITA-III limit on the all-flavor-sum diffuse
    ultra-high-energy neutrino flux and a combined limit from ANITA
    I-III, using the ANITA-III limit shown here and the published
    ANITA-II and ANITA-I limits~\cite{anita1,anita2}. The latest
    ultra-high-energy neutrino limits from the Auger~\cite{auger2015}
    and IceCube~\cite{icecube2017erratum} experiments, and two
    cosmogenic neutrino models~\cite{kotera,ahlers} are also
    shown. See Appendix~\ref{sec:limit} for details about the calculation of
    the limit. The table lists the ANITA-III effective area as a function of neutrino energy used to make the limit, not including analysis efficiency. } 
   \label{fig:limit}
 \end{figure}

 The limit (Fig.~\ref{fig:limit}) on the expected neutrino flux is calculated
 using a livetime of 17.4 days and a geometric mean of \texttt{icemc}-computed acceptance with an
 acceptance estimate from an independent MC simulation developed for ANITA, the
 analysis efficiency as a function of neutrino energy, and the appropriate 90\%
 Feldman-Cousins factor for the number of events detected and expected
 backgrounds. Further details are available in Appendix~\ref{sec:limit}. While Analysis~\textbf{A} would provide the best limit (as it finds no
 events), Analysis~\textbf{B} has the best expected sensitivity, so we use its result to
 set the limit.  The expected number of events for the Kotera maximum mixed-composition and maximum all-proton models, are $0.029\pm0.002$ and $0.17 \pm 0.01$, respectively. ANITA-III sets a 90\% CL integral flux limit on a pure $E_{\nu}^{-2}$ spectrum for $E_{\nu} \in [10^{18} \mathrm{eV},10^{21} \mathrm{eV}]$  of $E^2_{\nu} \Phi_{\nu} \leq  4.6 \times 10^{-7}~\mathrm{GeV}~\mathrm{cm}^{-2}~\mathrm{s}^{-1}~\mathrm{sr}^{-1}$.    

We also show a combined limit from ANITA I-III  where we have used the total
 number of events seen, total expected background, and the
 analysis-efficiency-weighted sum of previously-published effective
 volumes~\cite{anita2}.

\section{Discussion}

\newcommand{\evRA}{11.43} 
\newcommand{\evDEC}{16.3} 
\newcommand{\evMAJ}{5.0}
\newcommand{\evMIN}{1.0}
\newcommand{\evANG}{73.7}

 \begin{figure}
   \includegraphics[width=\columnwidth]{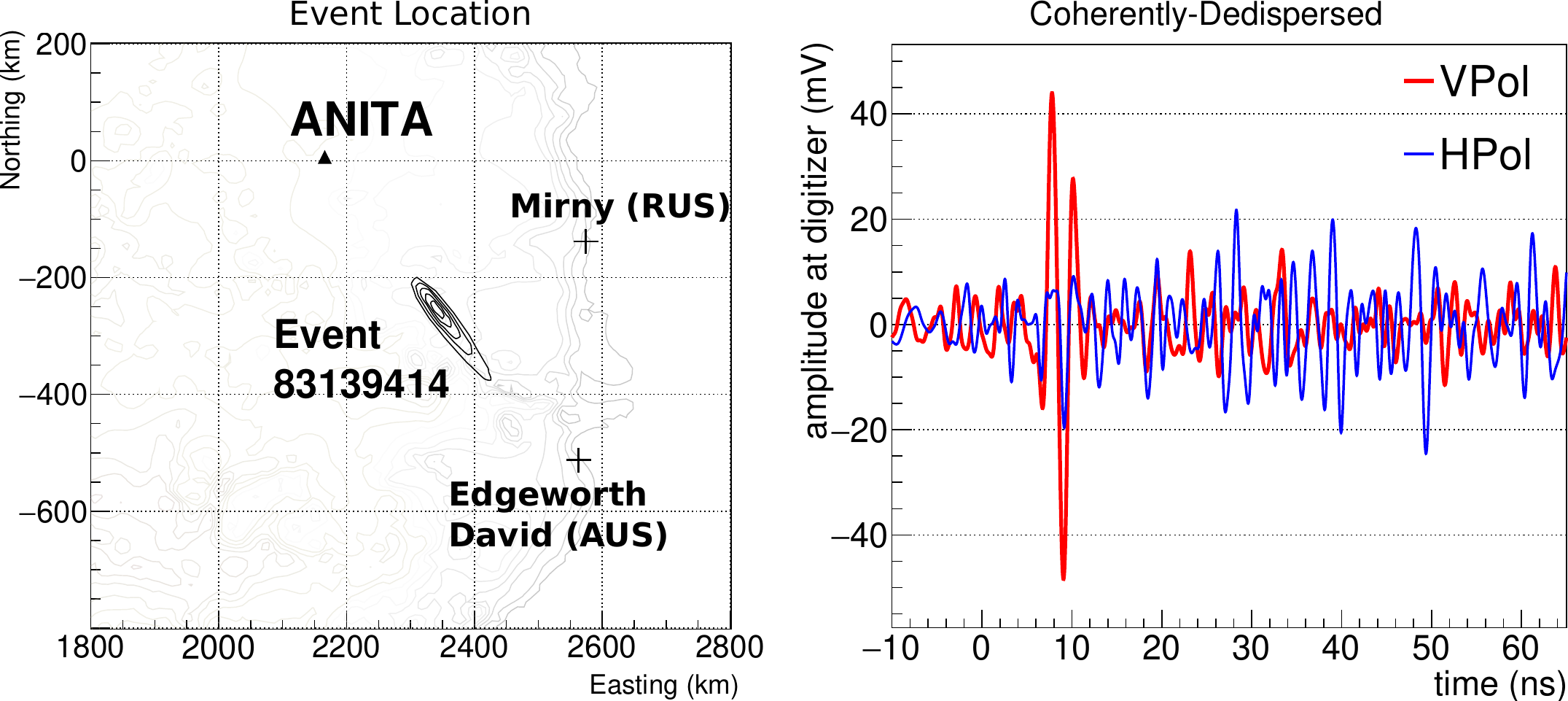} 
   \begin{tabular}{|l|l|} 
     \hline
     Estimated Event Location & 96.21$^\circ$ E, 68.57$^\circ$ S, 2319 m  \\
     Payload Location &  90.07$^\circ$ E, 70.26$^\circ$ S, 33643 m \\ 
     Minimum Energy & $10^{19}$ eV\\
     UTC Time & 2015-01-08 19:04:24.23740235\\
     Estimated Sky Position & \evRA~h, \evDEC$^\circ$ \\
     \hline
   \end{tabular}

   \caption{Event localization (left) and dedispersed coherently-averaged waveform (right) for vertically-polarized event 83139414. This event is in the Askaryan neutrino signal region in Analysis $\textbf{B}$ and a sub-threshold, isolated event in Analysis $\textbf{A}$.  The table below provides additional information about the event: the longitude, latitude, and ice depth at the estimated event location; the longitude, latitude, and altitude of payload at time of detection; and the minimum neutrino energy and mean sky position that could have produced the event according to the MC simulation. The black contours on the map represent 1-5 $\sigma$ regions for the event location. 
   	} 
     \label{fig:vpol_event} 
  \end{figure}

The isolated vertically-polarized event 83139414 from Analysis \textbf{B} (and just outside the signal region 
in \textbf{A}) is particularly intriguing. While consistent with
the pre-unblinding background estimate, our post-unblinding
interpretation is that the event is both unusually isolated and has a signal shape (Fig.~\ref{fig:vpol_event}) consistent with impulsive broadband emission.
There is no known human activity within 260~km. 

The polarization of the surviving event is 
consistent with expectations from neutrino
simulations, and the signal has no features
that make it easily identifiable as an anthropogenic signal ({\it e.g.} slow rise time, narrow bandwidth, double-pulse structure).
Including additional metrics such as these to distinguish neutrino-like signals from anthropogenic noise would 
reduce a post-unblinding
background estimate for this particular event by an order of magnitude.

The emission comes from a location on the continent consistent in ice depth and
elevation angle with simulated distribution of neutrinos.  The source location
of the emission is fully consistent with MC neutrinos simulations with the
ANITA-III flight path and recorded thresholds.
Simulations of neutrinos near the interaction location with the payload near
the detection position suggest a minimum possible neutrino energy of
$10^{19}$~eV.  Simulation may also be used to estimate the direction of a
neutrino producing radio emission compatible with the observed location and
polarization. The localization corresponding to the surviving event in equatorial
coordinates is well-approximated by an elliptical Gaussian centered at (RA,dec)
= (\evRA~h, \evDEC$^\circ$), with major and minor axis standard deviations of
\evMAJ$^\circ$ and \evMIN$^\circ$, respectively, and a position angle of
\evANG$^\circ$.

In summary, despite challenging EMI conditions ANITA-III has yielded a robust sample of radio-detected ultra-high-energy cosmic rays~\cite{a3tau}. While no compelling neutrino signal above background has been detected, the one remaining passing event in Analysis {\bf B} is rather unlike the parent anthropogenic population that comprises the typical background, but shares several important characteristics expected from neutrino events. 
\vspace{10pt}

\section{Acknowledgments} 
We would like to thank the National Aeronautics and Space Administration and the National Science Foundation.  We would especially like to thank the staff of the Columbia Scientific Balloon Facility and the logistical support staff enabling us to perform our work in Antarctica.  We are deeply indebted to those who dedicate their
careers to help make our science possible in such remote environments.
This work was supported by the Kavli Institute for Cosmological Physics at the University of Chicago.  Computing resources were provided by the University of Chicago Research Computing Center and the Ohio Supercomputing Center at The Ohio State University.
A. Connolly would like to thank the National Science Foundation
for their support through CAREER award 1255557. 
O. Banerjee and L. Cremonesi's work was also
supported by collaborative visits funded by the Cosmology and Astroparticle
Student and Postdoc Exchange Network (CASPEN).
The University College London group was also supported by the Leverhulme Trust. The Taiwan team is supported by Taiwan's Ministry of Science and Technology (MOST) under its Vanguard Program 106-2119-M-002-011.

\bibliography{main}

\appendix

\section{Analysis \textbf{A}}\label{sec:clustering}

\subsection{Filtering and interferometry}\label{sec:uc_filtering} 
The first step in Analysis~\textbf{A} is 
to mitigate man-made, narrow-frequency transmitter
contamination 
in each channel.
Such sources are referred here as continuous wave (CW).
We use the power spectrum to iteratively identify the highest-power frequency, which
constrains a trial 
sinusoidal fit in the 
waveform that we then subtract. 

\label{sec:interferometry}
We use these filtered waveforms and interferometry to generate 
the pointing map for each event, following the procedure in~\cite{interferometric}.  
We cross correlate pairs of nearby antennas (normalized by power), 
to extract time delays that in turn allow pointing. The procedure is applied to every point on the sky to make a map.
We consider the largest three isolated map peaks in each polarization as source candidates. We 
extract
azimuth ($\phi$) and elevation ($\theta$) using a quadratic
fit to a make higher-resolution map near the peaks.  

\subsection{Signal-like event selection} \label{sec:uc_observables} For each
candidate source, 
we compute a ``coherent waveform" by adding all channels with the proper delay. We also compute a ``dedispersed coherent waveform" using the measured instrument response and antenna response from that direction. 
We introduce an impulsivity measure, $\mathcal{I}$, by considering the average,
$A$, of the cumulative distribution of the fractional power contained within a
time before or after the peak of the Hilbert envelope (the magnitude of the analytic signal).  
We define $\mathcal{I}$ as 2$A$ - 1.
We choose the source candidate with largest $\mathcal{I}$ for further analysis. 

The first cut applied to the data is a requirement that the payload triggered on
the polarization of the most impulsive candidate. 
The second cut removes events with channel amplitudes
exceeding 1.0~V to remove digitizer glitches and saturating events.

The third cut targets payload-generated impulses (``payload blasts"), which originate from the payload and are impulsive, but are not the expected plane waves from distant sources. To remove these local events, we use a set of cuts that requires that all rings of antennas have comparable power,
and that the peak of the coherent average is compatible with the average of the individual waveforms that comprise it. 

The fourth cut aims to distinguish signal-like waveforms, with power to separate from thermal noise, any CW surviving the filtering process, and any payload blasts that survive at this stage. A Fisher
discriminant~\cite{fisher} is used with the following observables:  
1) normalized average correlations, 
2) the angular distance between reconstructed azimuth and mean trigger direction, 
3) Hilbert envelope peak of the dedispersed average, 
4) the difference between the average of individual waveform peaks and the coherent average peak, 
5) $\mathcal{I}$ for the dedispersed average,  
6) the difference of squares of the $\mathcal{I}$s of  dedispersed average and coherent average, 
7) the power-normalized difference between the peak of the dedispersed average and the coherent average, 
8) the full-waveform linear polarization fraction, 
9) the near-peak linear polarization fraction, and
10) the change in peak time between coherent and dedispersed averages. 

\begin{figure} 
  \includegraphics[width=\columnwidth]{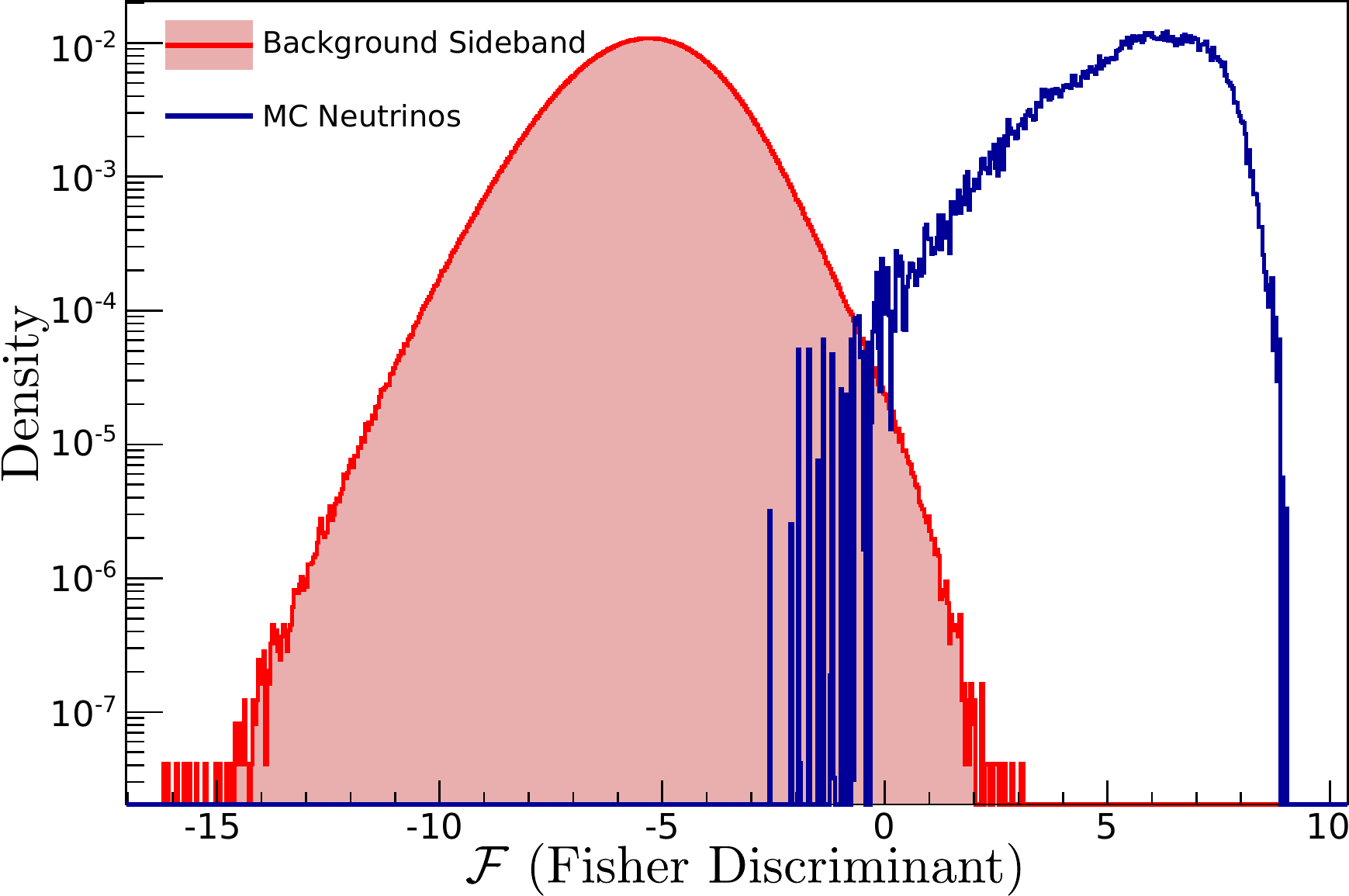} 
  \caption{Distributions of the multivariate score, $\mathcal{F}$, for weighted simulated neutrinos (blue) and the sideband reconstructing above the horizontal plane (red) we use as a background sample 
  to train the Fisher discriminant used in Analysis~\textbf{A}. }
  \label{fig:cosmin_fisher_distributions} 
\end{figure}

The discriminant is trained with the TMVA~\cite{tmva} framework using signal sample of simulated
neutrinos and a backgrand sample populated by events where the most
impulsive candidate reconstructs above the horizontal plane. 
The distributions of the multivariate analysis
discriminator value, $\mathcal{F}$, for the two samples is shown in
Fig.~\ref{fig:cosmin_fisher_distributions}.  The impulsivity variables dominate the power of the discriminant. 

\subsection{Isolation parameter (clustering)}

The final cut requires events to be
spatially isolated to separate neutrino-like events from anthropogenic events, which are likely to cluster with themselves. We localize each plane-wave
event ({\it i.e.} not thermal noise or payload blasts) reconstructed by ANITA to a particular direction in payload coordinates,
$(\phi_0,\theta_0)$ with an uncertainty we model as a bivariate
Gaussian distribution, $\mathcal{P}(\phi,\theta|\sigma_{\phi},\sigma_{\theta})$. 
We estimate $\sigma_{\phi}$ and~$ \sigma_{\theta}$ using the measured pointing resolution from calibration pulsers, with corrections for heading uncertainty and elevation angle. 

This distribution $\mathcal{P}(\phi,\theta)$ projected onto Antarctica
is called $\mathcal{P}'(\mathrm{E},\mathrm{N})$.  
We exclude portions of $\mathcal{P}$ that miss the ground and renormalize the distribution. We exclude events with less than 0.1\% of $\mathcal{P}$ on the
continent or with elevation more than $40^\circ$ below the horizontal plane. We then compare $\mathcal{P}'$, which is an event's localization distribution, to that of other events.

The  
well-known Bhattacharyya Coefficient~\cite{bhattacharyya}, 
$ BC(p_i(\vec{x}), p_j(\vec{x})) = \int \sqrt{ p_i p_j} d \vec{x} $,  quantifies the similarity of two distributions.
We extend this general idea to compute a global
overlap, $\mathcal{O}$, of each event, $i$, with all other events of interest, $j$, as: 

\begin{equation}
  \mathcal{O} = \sum_{j}  \int_{Antarctica} \sqrt{ w_i w_j \mathcal{P}'_j  \mathcal{P}'_i} , 
\label{eqn:logO} 
\end{equation}
where $w$ is a weighting factor.  An
event originating near many other events of interest will have a large
$\mathcal{O}$, while an isolated event has $\mathcal{O}\approx 0$.
The variable $\mathcal{O}$ spans many orders of magnitude, so it is convenient to work with $-\log\mathcal{O}$ as an isolation parameter. 

The sample of events used for clustering is defined using the Fisher score
$\mathcal{F}$.
We find that there is a change in the shape of the background sample
distribution along the Fisher score at $\mathcal{F}>2$, so we include all
events above that score in the clustering sample. Between $\mathcal{F}>2$ and
$\mathcal{F}<3.2$, where we ultimately set our final cut on $\mathcal{F}$ that
defines the signal region, we assign a weight between 0 and 1 in the clustering
sample based on the $\mathcal{F}$ value, with a weight of 1 for events at
$\mathcal{F}=3.2$ and a weight of 0 for events at $\mathcal{F}=2$.  Above
$\mathcal{F}=3.2$, we assign a weight of 1 for each event.

\subsection{Setting final cut values}\label{sec:uc_sett-fish-discr}
 
The final cut thresholds for $\mathcal{F}$ and $-\log \mathcal{O}$ are set to optimize the sensitivity of the Askaryan neutrino search on the Kotera flux model. 
We compute the analysis efficiency using simulated
neutrinos generated according to a Kotera flux model, and passing
them through the analysis. For the isolation cut, each neutrino is individually
evaluated against the entire source map.
We estimate a 5\% systematic uncertainty on analysis efficiency
based on a comparison with the calibration pulsers deployed on the ice. 

The two sources of background, leakage from events cut by the Fisher discriminant ($\mathcal{B}_{\mathcal{F}}$) and leakage from isolated anthropogenic signal-like events ($\mathcal{B}_{\mathcal{O}}$), are each estimated
from background regions, as in the on-off problem.  We estimate $\mathcal{B}_{\mathcal{F}}$ from the above-horizontal background region depicted in
Fig.~\ref{fig:cosmin_fisher_distributions}.  We calculate
the leakage in the background region, and 
multiply by the ratio of events in the background to those below the horizon.

We lack a true signal-free background region to estimate $\mathcal{B}_{\mathcal{O}}$. We
instead use the sub-threshold region $\mathcal{F} \in (2,2.8)$, where the upper
threshold remains below any reasonable cut in $\mathcal{F}$ to avoid accidental
unblinding, and use the distribution of events in this region in $-\log\mathcal{O}$ to estimate the background in the region of interest. 
We consider events in both polarizations but exclude large clusters. 

We use a profile-likelihood method~\cite{Rolke} implemented with
RooStats~\cite{RooStats} 
to optimize the final cut on
$\mathcal{F}$ and $-\log\mathcal{O}$. Using this model, we find the search to be most
sensitive with a cut on $\mathcal{F} > 3.2$ and $-\log \mathcal{O} > 12$. 

Each polarization is expected to contribute equally to background. 
The total estimated background from all sources for this search in either the
horizontal or vertical polarization region is $0.78^{+0.60}_{-0.39}$. 

 \begin{figure} 
   \includegraphics[width=\columnwidth]{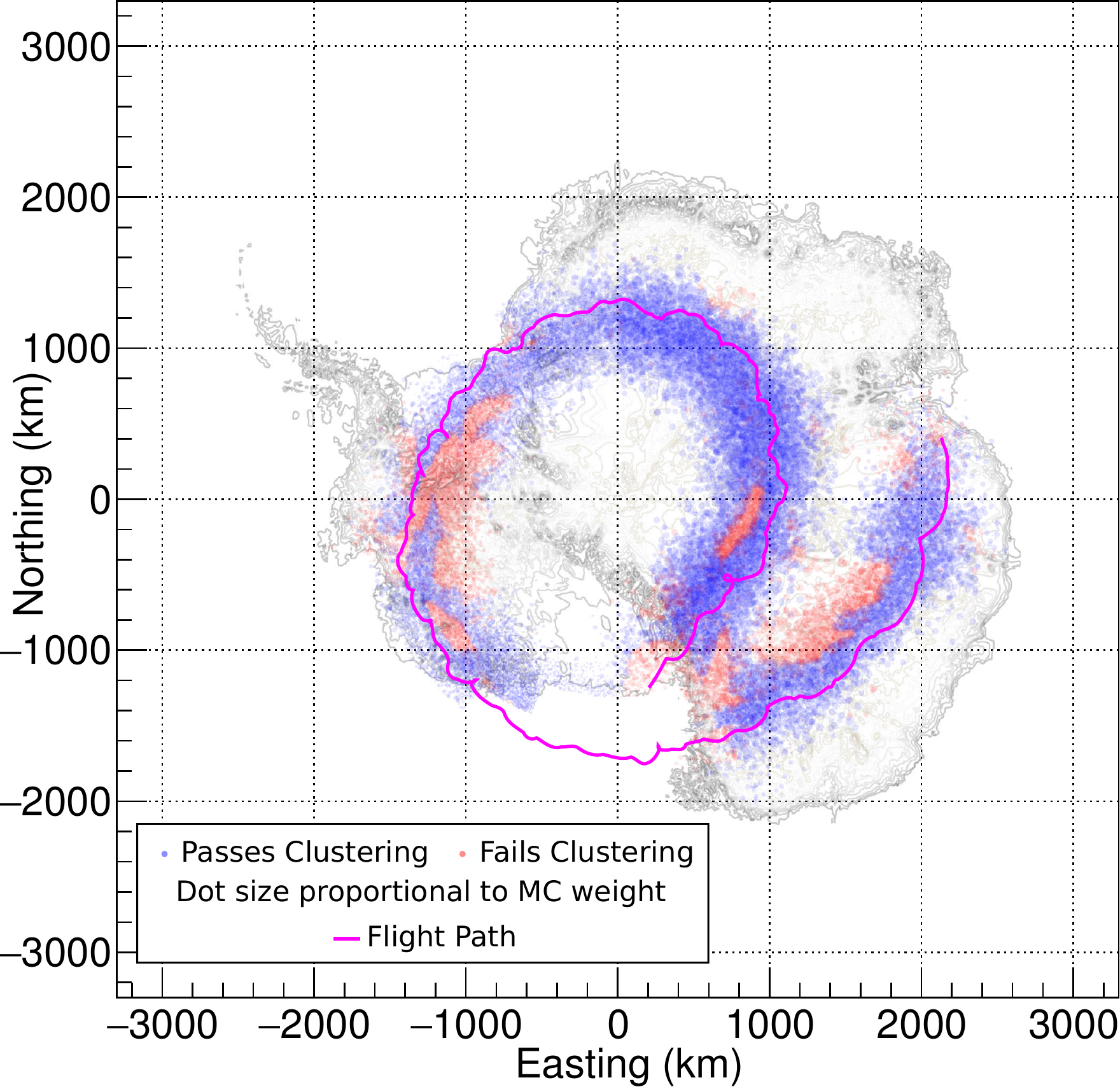}

   \caption{
     Clustering efficiency on simulated neutrinos in Analysis \textbf{A}.   	
     Each dot
   represents a MC neutrino event passing pre-clustering cuts, with area
   scaled to MC weight. Blue points indicate events that pass clustering
    after removing likely extensive air showers; 
   	red points are those that fail clustering. } 

   \label{fig:cluster_eff} 
 \end{figure}

Fig.~\ref{fig:cluster_eff}
shows the clustering efficiency of MC neutrino events surviving pre-clustering cuts as a
function of position on the continent. The analysis efficiency is 88\% pre-clustering and 72\% after all cuts are applied. 

\subsection{Results}\label{sec:uc_results}

\begin{table}

  \begin{tabular}{c||c|c|c} 
Cut                     & MC $\nu$    & VPol        &   HPol               \\ 
                        & Efficiency  & Remaining   &  Remaining           \\
\hline                                              
\hline                                              
None                    & 1.00        & 38,274,132  & 36,700,502           \\
Trigger Correct Pol.    & 0.98        & 20,599,991  & 18,825,981           \\
Peak $<$ 1000            & 0.96        & 20,565,939  & 18,811,772           \\
Payload Blast           & 0.95        & 16,474,185  & 15,655,493           \\
Elevation $<$ 0         & 0.95        &  3,821,760  & 3,802,329            \\
\hline                                              
$\mathcal{F} > 3.2$     & 0.88        &169,824      & 311,795              \\
\hline                                              
$-\log\mathcal{O} > 12$ & 0.72        & 0           & 22                   \\
\hline
\end{tabular}
\caption{The effect of each cut on horizontally-polarized and vertically-polarized science triggers as well as the 
	efficiency of each cut on MC-generated neutrinos for Analysis~\textbf{A}.  The bottom
    row in the table shows the remaining events
    in the horizontally and vertically-polarized signal regions, as 
    well as the total analysis efficiency after all cuts are 
    applied.} 
\label{tbl:uchicago_cut_table} 
\end{table}

After unblinding, zero events were found in the vertically-polarized signal
region.  This corresponds to a 90\% upper limit on triggered neutrino
events of 1.37 using the prescription outlined before and 1.67 using a standard
Feldman-Cousins method integrated over systematics. While not in the signal region, there was one isolated,
vertically-polarized event (83139414, $\mathcal{F}=3.1$) below the final cut threshold on
$\mathcal{F}$, which is a broadband, impulsive, isolated event. This is the
same event found in Analysis~\textbf{B} and depicted in
Fig.~\ref{fig:vpol_event}. 

In the horizontally-polarized channel, 22 events were found in the signal
region, 21 of which have a waveform shape and polarization consistent with
geomagnetic emission from Extensive Air Showers (EAS) created by cosmic rays. The other was inconsistent with the EAS hypothesis (in both shape and polarization), but is consistent with our background estimate. Eighteen of these events were also
found in a dedicated air shower search~\cite{BRotterThesis} that used an
independent method relying on cross correlation with a waveform template from a cosmic-ray simulation. 

Table~\ref{tbl:uchicago_cut_table} shows the effect of each cut on the data and on simulated neutrinos. 

\newcommand{\coherent}{\mathcal{C}}
\newcommand{\dedispersed}{\mathcal{D}}
\newcommand{\phisector}[1]{$\Phi{}$-sector#1}
\newcommand{\hical}[1]{HiCal#1}
\newcommand{\peakHilbert}{\mathcal{H}}
\newcommand{\impulsivityMeasure}{\mathcal{I}}
\newcommand{\fracPowerWindowGradient}{\mathcal{G}}
\newcommand{\mapPeak}{\mathcal{P}}
\newcommand{\peakTheta}{\theta_{\mapPeak}}
\newcommand{\peakPhi}{\phi_{\mapPeak}}
\newcommand{\fisherDiscriminant}{\mathcal{F}}
\newcommand{\fisherThreshold}{\fisherDiscriminant{}_{\text{threshold}}}
\newcommand{\fisherThresholdValue}{5.8} 
\newcommand{\chisquarePerNdf}{\chi^{2}/ndf}

\newcommand{\llThresh}{\Lambda} 
\newcommand{\dThresh}{\lambda} 

\newcommand{\eventEventFitInput}{\vec{s}}
\newcommand{\eventEventLikelihood}{L(\eventEventFitInput)_{ij}}
\newcommand{\eventEventLogLikelihood}{-2 \log(\eventEventLikelihood)}
\newcommand{\eventBaseLogLikelihood}{{-2 \log(L)}_{ib}}
\renewcommand{\todo}[1]{\textcolor{red}{TODO: #1}}
\newcommand{\analysisA}{Analysis \textbf{A}}
\newcommand{\analysisB}{Analysis \textbf{B}}

\newcommand{\updatedClusteringBackground}{$0.50^{+0.50}_{-0.25}$}
\newcommand{\nonImpulsiveBackground}{$0.16 \pm 0.01$}

\section{Analysis B}\label{sec:clustering2}

\subsection{Blinding}\label{sec:blinding}
\analysisB{}, like \analysisA{}, did not look in the hidden signal region until all cuts were set.
Additionally, a small number (unknown to the analyst) of vertically-polarized calibration pulser events were inserted randomly throughout the data.
These were removed after opening the blinded region.
No horizontally-polarized events were inserted.

\subsection{Filtering, reconstruction and quality cuts}\label{sec:narrow-band-cw}
\analysisB{} uses the same filtering method as \analysisA{}, and a similar reconstruction method (Section~\ref{sec:uc_filtering}).
Unlike \analysisA{}, we use the source candidate with the largest map peak value, $\mapPeak$, to define the event direction, $\peakPhi, \peakTheta$, and primary polarization.

We restrict the analysis sample to events with $\peakTheta < 0$, and remove ground station calibration pulses from the data set using subsecond timing information.

We first remove the small subsample of data that is poorly recorded or reconstructed by requiring good GPS data, sufficient samples in the digitized waveforms, and consistent pointing $(\peakTheta, \peakPhi)$ at each stage of the directional reconstruction (Section~\ref{sec:interferometry}).

We then use a series of cuts to remove payload blasts. We remove events that do not have similar power 
or peak amplitude in the top and bottom antenna rings.
Events where the Hilbert envelope peak, $\peakHilbert$, does not scale as expected with the top and bottom antenna ring amplitudes or map peak, $\mapPeak$, are also removed.
The effect of these cuts on the data and a Kotera flux of simulated neutrinos is given by the ``quality cuts'' row in Table~\ref{tab:ucla_cuts}.

\subsection{Impulsivity cuts}\label{sec:thermal-weak-cw}
Like \analysisA{}, Analysis~\textbf{B} trains a Fisher discriminant, $\fisherDiscriminant$, to separate simulated neutrino events from a non-impulsive background region, triggered events with ($\peakTheta > 0$).

\begin{figure}
  \includegraphics[width=\columnwidth]{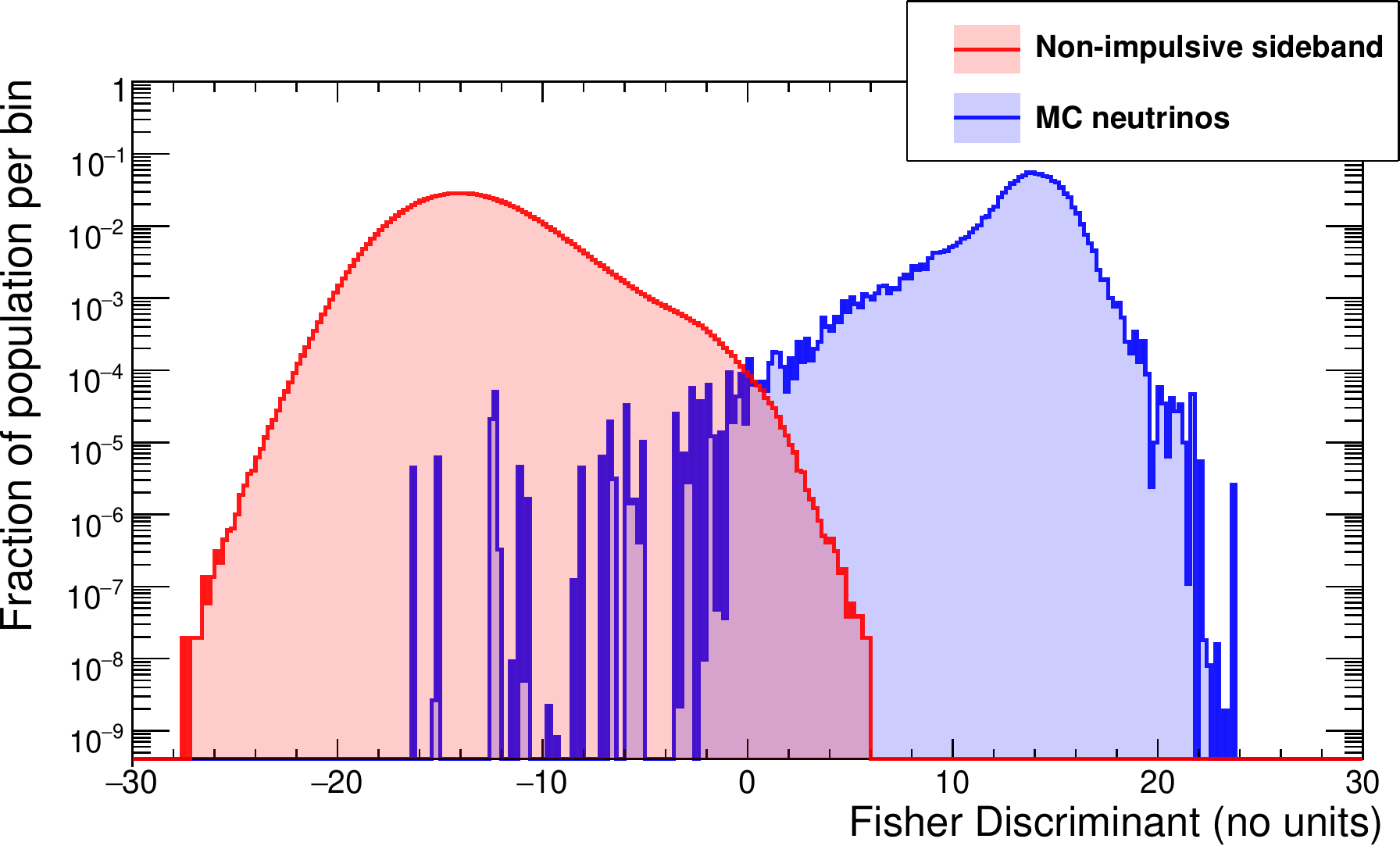} 
  \caption{The distribution of the \analysisB{} Fisher discriminant, $\fisherDiscriminant$, for the non-impulsive background region and simulated neutrinos.
  }\label{fig:ucla_fisher} 
\end{figure}

The Fisher discriminant $\fisherDiscriminant$ is a weighted sum of 7 variables: the map peak, $\mapPeak$; the Hilbert envelope peak of the coherently averaged waveform before ($\peakHilbert_{\coherent}$) and after ($\peakHilbert_{\dedispersed}$) dedispersion; and \analysisA{}'s impulsivity measure before ($\impulsivityMeasure_{\coherent}$) and after ($\impulsivityMeasure_{\dedispersed}$) dedispersion.
\analysisB{} uses a new variable, the power window gradient, $\fracPowerWindowGradient$, defined as the average difference between the five smallest time windows containing 10\%, 20\%, 30\%, 40\%, and 50\% of the total power.
The final two components of $\fisherDiscriminant$ are $\fracPowerWindowGradient_{\dedispersed}$, and $\fracPowerWindowGradient_{\coherent}/\fracPowerWindowGradient_{\dedispersed}$, where the subscripts $\coherent$ and $\dedispersed$ correspond to the coherently averaged waveform before and after dedispersion respectively.
The distributions of $\fisherDiscriminant$ are shown in Fig.~\ref{fig:ucla_fisher}.

  \begin{table}
    \begin{tabular}{c||r|r|r}
      Cut & \multicolumn{1}{c}{MC $\nu$}  & \multicolumn{1}{|c}{VPol}      & \multicolumn{1}{|c}{HPol}      \\
      {}  & \multicolumn{1}{c}{Efficiency} & \multicolumn{1}{|c}{Remaining} & \multicolumn{1}{|c}{Remaining} \\
      \hline
      \hline
      Not ground pulser & 1.00                & 8,888,370                 & 10,767,799              \\
      Close to MC truth & 0.99                & {}                      & {}                    \\
      Quality cuts      & 0.97                & 8,125,293                 &  9,940,345              \\
      Pass thermal cut  & 0.96                &  242,957                 &   361,383              \\
      Not HiCal         & 0.95                &  242,941                 &   360,604              \\
      \hline
      Clustering        & 0.84                &  1                      &   25                  \\
      \hline
    \end{tabular}
    \caption{Events passing cuts in sequence in Analysis~\textbf{B}, similar to Table~\ref{tbl:uchicago_cut_table} for Analysis~\textbf{A}.
      The final total of vertically-polarized events is given after removing inserted events.
    }
    \label{tab:ucla_cuts}
  \end{table}

To remove non-impulsive events from the analysis sample we require $\fisherDiscriminant > \fisherThresholdValue$.
We model the tail of the non-impulsive background region as an exponential, fitting over a range of tail values.
The cut was set such that we expect fewer than 0.5 events in the search region, by multiplying the number of expected events above the cut in the background region by the ratio of events in the background region to the search region.
We treat each polarization symmetrically, and estimate a non-impulsive background of \nonImpulsiveBackground{} events per polarization.

We remove calibration pulser events from \hical{}~\cite{hical}, a balloon-mounted calibration pulser, with an azimuth pointing cut $|\phi_{hc} - \peakPhi | < \SI{5}{\degree}$, where $\phi_{hc}$ is the azimuth of \hical{} in ANITA payload coordinates.

The effect of these cuts for events passing quality cuts is given in Table~\ref{tab:ucla_cuts}.

\subsection{Clustering algorithm}\label{sec:surface-clustering}
                         
For each event, we project the best direction hypothesis along payload coordinates $(\peakPhi, \peakTheta)$ onto a model of the Antarctic surface~\cite{bedmap2}.
To group surviving events together we use a clustering algorithm with two metrics: Euclidean separation on the surface and an angular separation at the payload.
Events $i$ and $j$ belong to the same cluster if their Euclidean separation, $d_{ij} \equiv | \vec{x}_{i} - \vec{x}_{j} |  < \dThresh$, where $\dThresh$ is a threshold distance.
For the angular clustering we perform a fit for each pair of events $i$, $j$, to minimize a ``same-source likelihood'', $\eventEventLikelihood$, given by
\begin{align}
  \label{eq:event-event-clustering-log-likelihood}
  \eventEventLogLikelihood &= {\left( \frac{ \theta_{i} - \theta_{si}}{\sigma_{\theta i}} \right)}^{2} + {\left( \frac{ \phi_{i} - \phi_{si}}{\sigma_{\phi i}} \right)}^{2} \nonumber \\
                           &+ {\left( \frac{ \theta_{j} - \theta_{sj}}{\sigma_{\theta j}} \right)}^{2} + {\left( \frac{ \phi_{j} - \phi_{sj}}{\sigma_{\phi j}} \right)}^{2} \qquad .
\end{align}
The fit varies a position on the surface, $\eventEventFitInput$, which appears at payload coordinates $(\phi_{si}, \theta_{si})$ for event $i$.
The event $i$ angular resolution, $\sigma_{\phi i}$ and $\sigma_{\theta i} $, is estimated from the resolution of calibration pulse events with the same SNR.
Events $i, j$ are considered clustered if $\eventEventLogLikelihood < \llThresh$, where $\llThresh$ is an angular threshold.
\begin{figure}[h]
  \includegraphics[width=\columnwidth]{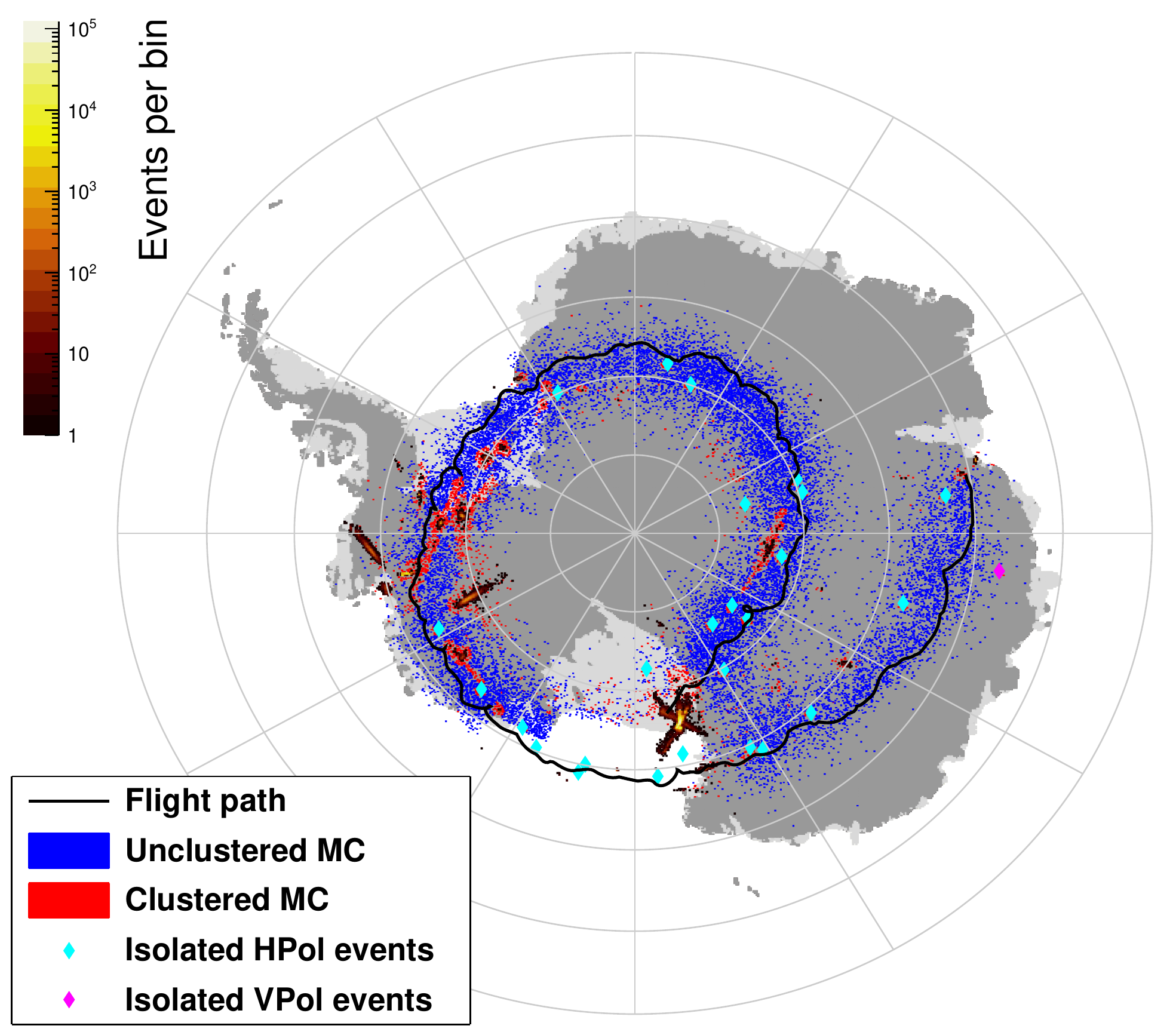}
  \caption{
    The histogram shows all events passing pre-clustering cuts in \analysisB{} projected onto the continent along with a subset of the simulated neutrinos passing (blue) and failing (red) clustering cuts.
    The positions of the 25 isolated horizontally-polarized (cyan) and vertically-polarized (magenta) events are also shown.
  }\label{fig:ucla_event_projection}
\end{figure}

\subsection{Base association}\label{sec:base-association}
We obtained a list of bases active in 2014--2015 by correspondence with a variety of Antarctic national programs.
We exclude regions surrounding each base using the same metrics as the event-event clustering (Section~\ref{sec:surface-clustering}).
A cluster of events is associated with a base, $b$, if their separation, $d_{ib} = | \vec{x}_{i} - \vec{x}_{b} | < \dThresh$.
A cluster is also associated with a base if the angular separation, $\eventBaseLogLikelihood$, of any event $i$ satisfies:
\begin{equation}
  \label{eq:event-base-clustering-angular-likelihood}
  \eventBaseLogLikelihood = {\left( \frac{ \theta_{i} - \theta_{b}}{\sigma_{\theta i}} \right)}^{2} + {\left( \frac{ \phi_{i} - \phi_{b}}{\sigma_{\phi i}} \right)}^{2} < \llThresh ,
\end{equation}
where $(\phi_{b}, \theta_{b})$ is the location of base $b$ in payload coordinates for event $i$.
\subsection{Setting clustering thresholds}\label{sec:sett-clust-cuts}
We estimate an anthropogenic background from the possible incompleteness of our base list using an $ABCD$ method, where 
$A$ is the number of small multiplets associated with known bases, $B$ is the number of small multiplets \emph{not} associated with known bases, $C$ is the number of singlets associated with known bases, and $D$ is the number of singlets \emph{not} associated with known bases (signal region).
$D$, is estimated as:
\begin{equation}
  \label{eq:abcd}
  D = C \times \frac{B}{A} . 
\end{equation}

The values of $A, B, C$ and $D$ depend on $\dThresh$ and $\llThresh$.
They also depend on the location of the inserted events (Section~\ref{sec:blinding}).
If the clustering is efficient, most inserted events will end up in category $D$.
However, if any end up in $A, B$ or $C$, it will change the background estimate.
To set the clustering thresholds we make an initial background estimate (to be revised in the case that any inserted events are found in categories $A, B$ or $C$ after opening the box) by defining a ``small multiplet'' as containing 2--4 events, and without modelling any uncertainty on the size of the known activity background region $\tau_{abcd} = A/B$.

Similar to \analysisA{} (Section~\ref{sec:uc_sett-fish-discr}), we use a profile-likelihood method~\cite{Rolke} implemented with RooStats~\cite{RooStats} to choose $\dThresh$ and $\llThresh$ to maximize sensitivity to a Kotera flux of neutrinos.
We perform a scan through $1 \leq \llThresh \leq 4000$ with $\dThresh = $ 30~km, 40~km, 50~km, finding $\dThresh = 40$~km and $\llThresh = 4$ give the best sensitivity.
Applied to simulation, the neutrino clustering efficiency is 89\%, when applied after the impulsivity cuts the total analysis efficiency is 84\%.  We estimate a 3\% systematic uncertainty on the efficiency based on a comparison with calibration pulsers.
The provisional background estimate from clustering alone (vertical and horizontal polarizations together) is 0.67 events.

\subsection{\analysisB{} results}\label{sec:analysisb-results}

After unblinding, \analysisB{} finds 25 horizontally-polarized events and 13 vertically-polarized events, of which 12 were inserted as a form of blinding (Section~\ref{sec:blinding}).
The distribution of impulsive events from the continent with simulated neutrinos passing and failing clustering cuts is shown in Fig.~\ref{fig:ucla_event_projection}.
The 12 inserted events that passed the analysis have been removed.

There were 14 inserted vertically-polarized events in total, giving an inserted event efficiency of 85\%, which is consistent with the efficiency estimate from the simulation.
The two inserted events that failed the analysis clustered with one-another in a non-base multiplet of size two, falling into category $B$ in Equation~\ref{eq:abcd}.
After removing these events we re-estimate the anthropogenic background, accounting for two additional sources of uncertainty not considered in Section~\ref{sec:sett-clust-cuts}.
We allow the known-base background region ratio, $\tau_{\text{abcd}} = A / B$ (from Equation~\ref{eq:abcd}), to vary by changing the upper limit on small multiplets (Section~\ref{sec:base-association}) from 4 to 9.
Additionally, we construct a distribution for $D$ by drawing from three Poisson distributions with means $A$, $B$, and $C$.
The updated background estimate was divided equally between horizontal and vertical polarizations by setting  $\tau_{\text{abcd}}(\text{VPol}) = \tau_{\text{abcd}}(\text{HPol}) = 2\tau_{\text{abcd}}$.
This leads to an anthropogenic background estimate of \updatedClusteringBackground{}.

Combining this estimate with the non-impulsive background estimate (Section~\ref{sec:thermal-weak-cw}), the final result of \analysisB{} is 25 horizontally-polarized events, all consistent with emission from cosmic ray air showers, on an expected background of \combinedBackgroundEstimates{}, and one vertically-polarized event in the Askaryan neutrino signal region on an expected background of \combinedBackgroundEstimates{}.
Details of the events are discussed in Section~\ref{sec:results}.

\section{Analysis C} 
\label{sec:binned}
Analysis {\bf C} imposes geographically-dependent 
selection criteria that are designed to be optimal
for  the local noise environment.  This approach has
the potential to recover sensitivity in regions of ice
rejected in clustering-based searches, and reach lower
thresholds in regions of ice that are free of anthropogenic backgrounds. 

\subsection{\bf{Cuts based on Circular Polarization (CP) and the Linear Discriminant (LD) cut}}  
\begin{figure}
\centering
\includegraphics[width=0.9\columnwidth]{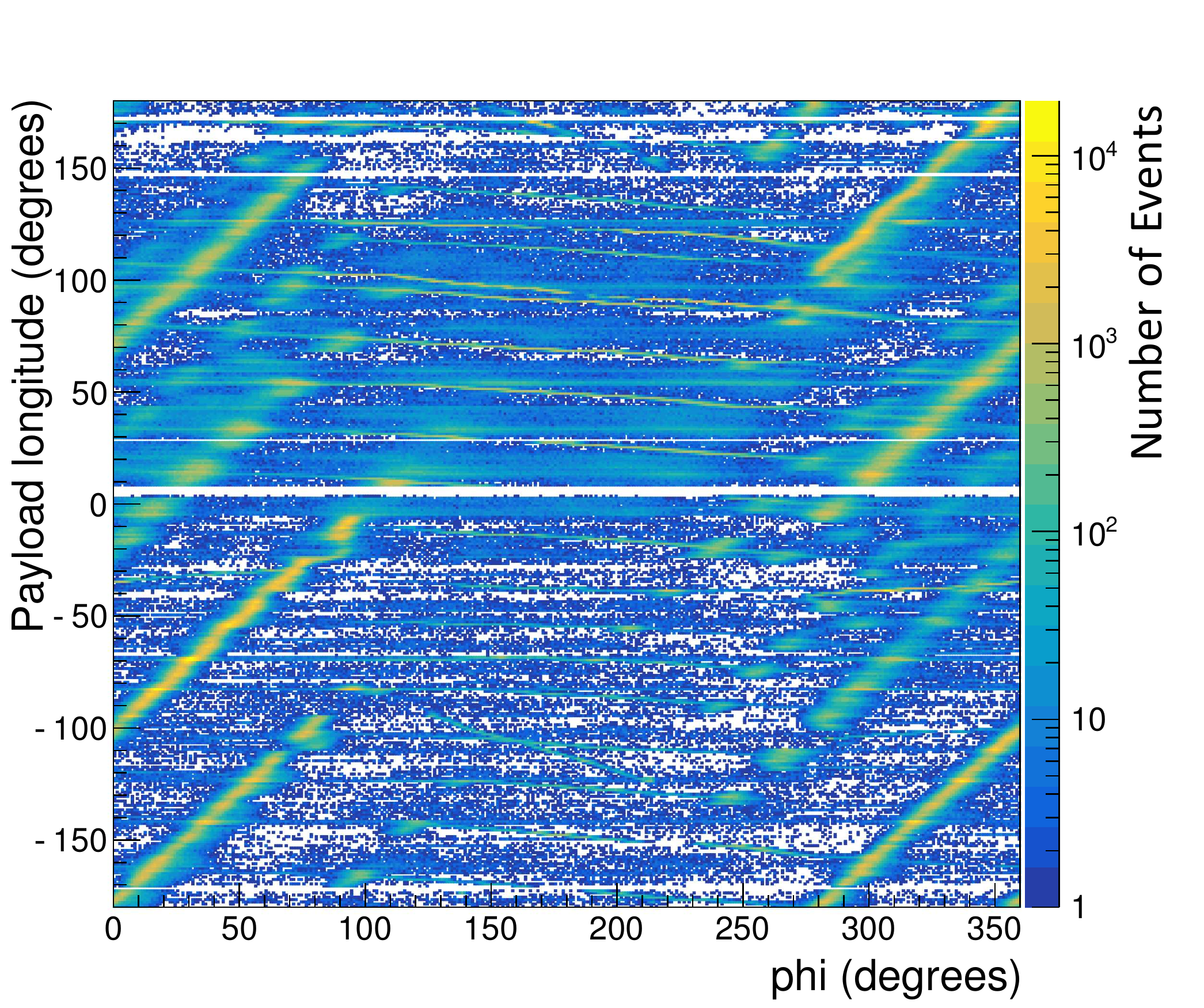}
\caption{The longitude of the ANITA-III payload as a function of reconstructed angle in azimuth, corrected for payload heading, derived from the cross correlation of left-circularly polarized (LCP) waveforms. CW signals  from geosynchronous satellites appear as an excess of events along lines with a slope of unity. ``Stripes'' are faint near $\phi=180^{\circ}$, where the reconstructed location is towards South Pole.}
\label{fig:stripes}
\end{figure}

Analysis {\bf C} is unique in performing event reconstruction in circular polarization. 
Being linearly polarized, the Cherenkov emission from neutrinos are expected to show strong cross correlation in both circular polarizations for relative
delays corresponding to the incoming radio-frequency direction, unlike
thermal noise and satellite CW signals. ANITA-III observed CW emission from geosynchronous satellites with a strong left circular polarization (LCP) component.
In Fig.~\ref{fig:stripes},  we plot the ANITA payload longitude on the vertical axis and the  reconstruction 
in azimuth  of events in LCP on the horizontal axis, with color representing the number of events.
Distinct over-densities or ``stripes'' can be seen with the slope expected for geosynchronous orbits. 

To select linearly-polarized signals, based on an optimization for the best expected limit at an early stage of the analysis, we remove events with a cross-correlation peak in LCP greater than 46$^{\circ}$ from that in right circular polarization (RCP), and with the peak-normalized cross-correlation value in either circular polarization below 0.015. 
These cuts predominantly remove thermal noise rather than satellite interference. We also reject events whose LCP reconstruction map peaks in a direction corresponding to one of the stripes seen in Fig.~\ref{fig:stripes}, if a geostationary satellite at a given azimuthal angle would be visible to ANITA given the payload latitude.

\begin{figure}
\centering
  \includegraphics[width=0.9\columnwidth]{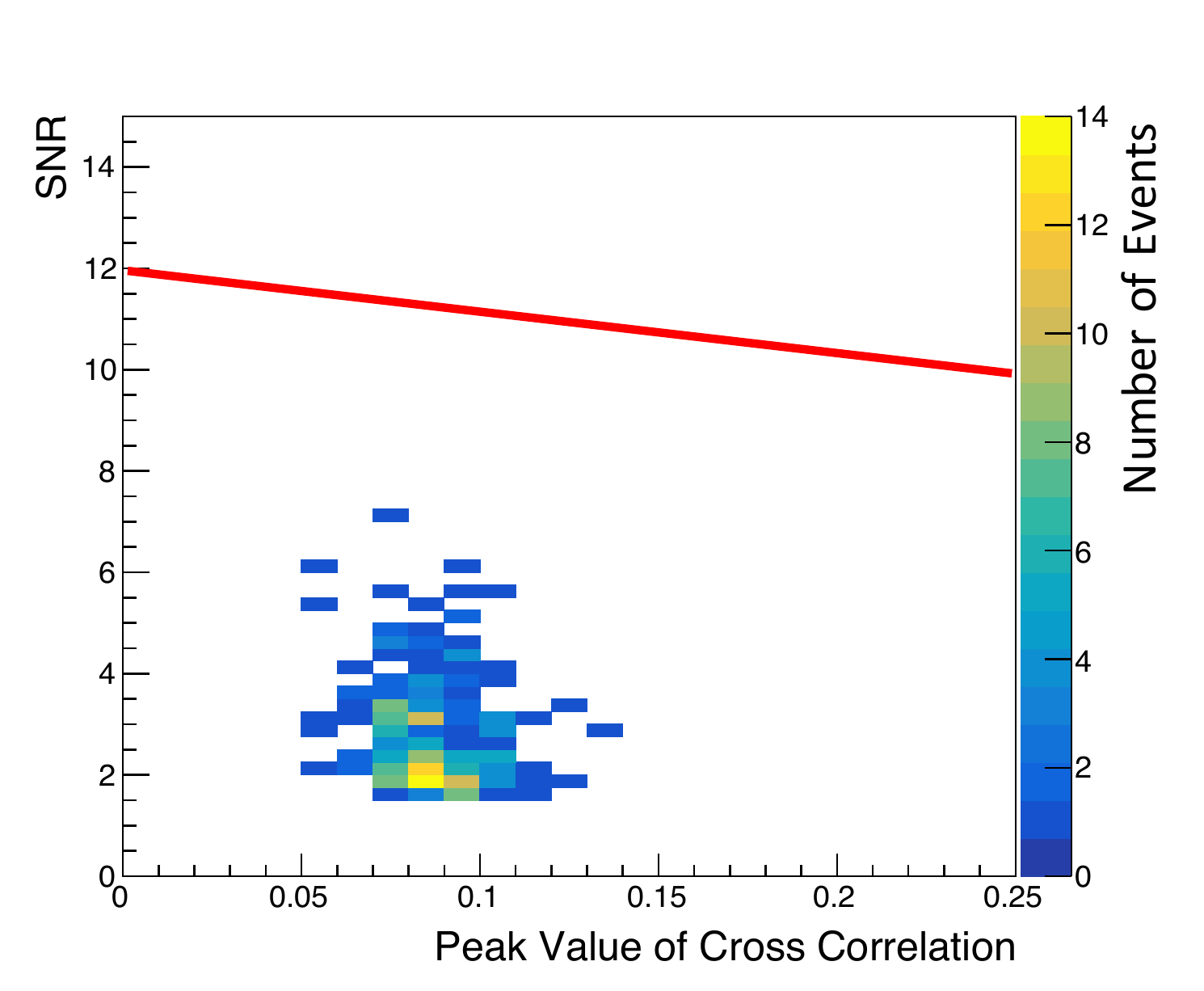}\hfill
\caption{The voltage SNR of the coherently summed waveform as a function of peak cross-correlation value for events in the 10\% dataset of ANITA-III used in Analysis~\textbf{C}, reconstructing to HEALPix bin 2970, in the vertical polarization analysis. The red line shows the LD cut for this bin.}
\label{fig:ld2d}
\end{figure}

Whereas Analyses {\bf  A} and {\bf B} use
selection criteria with a focus on measures of impulsivity of the signals, one way in which Analysis {\bf C} is complementary is its focus on peak cross-correlation values. Fig.~\ref{fig:ld2d} shows a two-dimensional distribution of the voltage SNR 
of the coherently summed waveform plotted against the peak value of the cross-correlation coefficient of events reconstructing to one of
the geographic bins
in the ANITA-III dataset. All events below the red line in the plot are cut.  A common slope for the cut line 
was used across all bins at -6 (unitless) as in 
Figure~\ref{fig:ld2d}, since at an early stage of the analysis
it was found to produce the best expected limit. A unique and optimum placement of the LD cut is calculated for each bin, as described in Section~\ref{optimization}. 

\begin{figure}
\centering
\includegraphics[width=0.9\columnwidth]{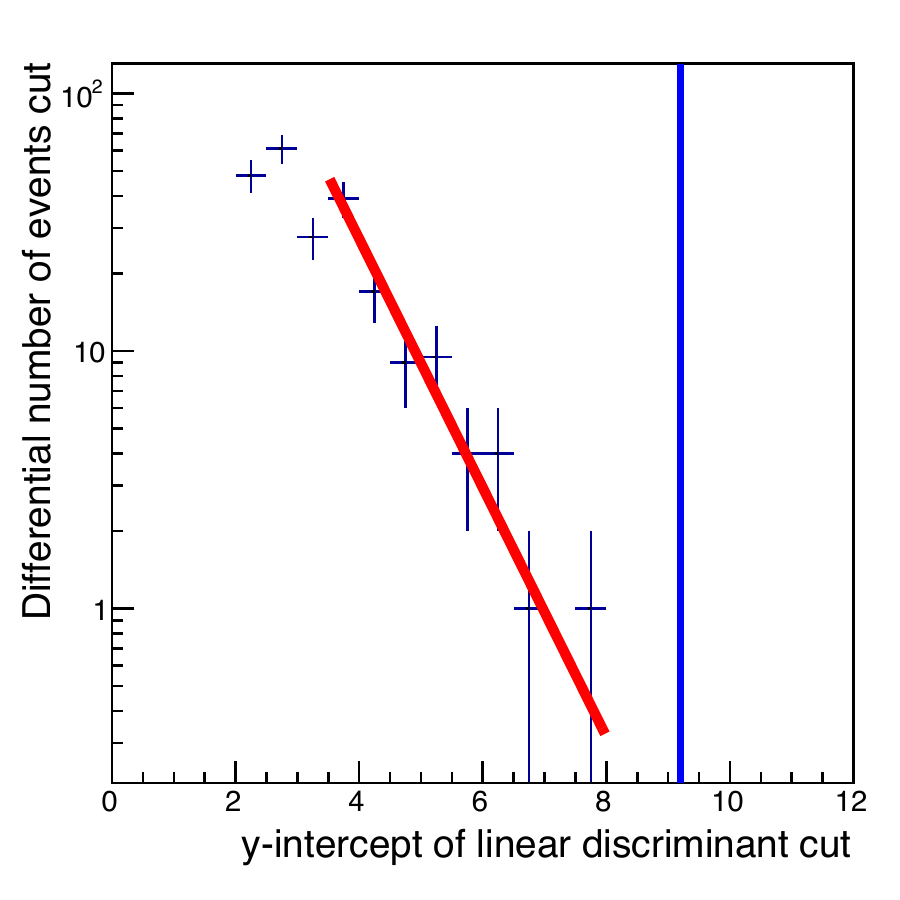}
\caption{The distribution of events as a function 
of position of the linear discriminant cut 
from the 10\% dataset in Analysis~\textbf{C} for bin 2970 in the VPol
channel of the analysis.  The blue vertical line shows the position of the cut in this bin found by the optimization procedure.}
\label{fig:diffdist}
\end{figure}

\begin{figure*}
\centering
\includegraphics[width=1.6\columnwidth]{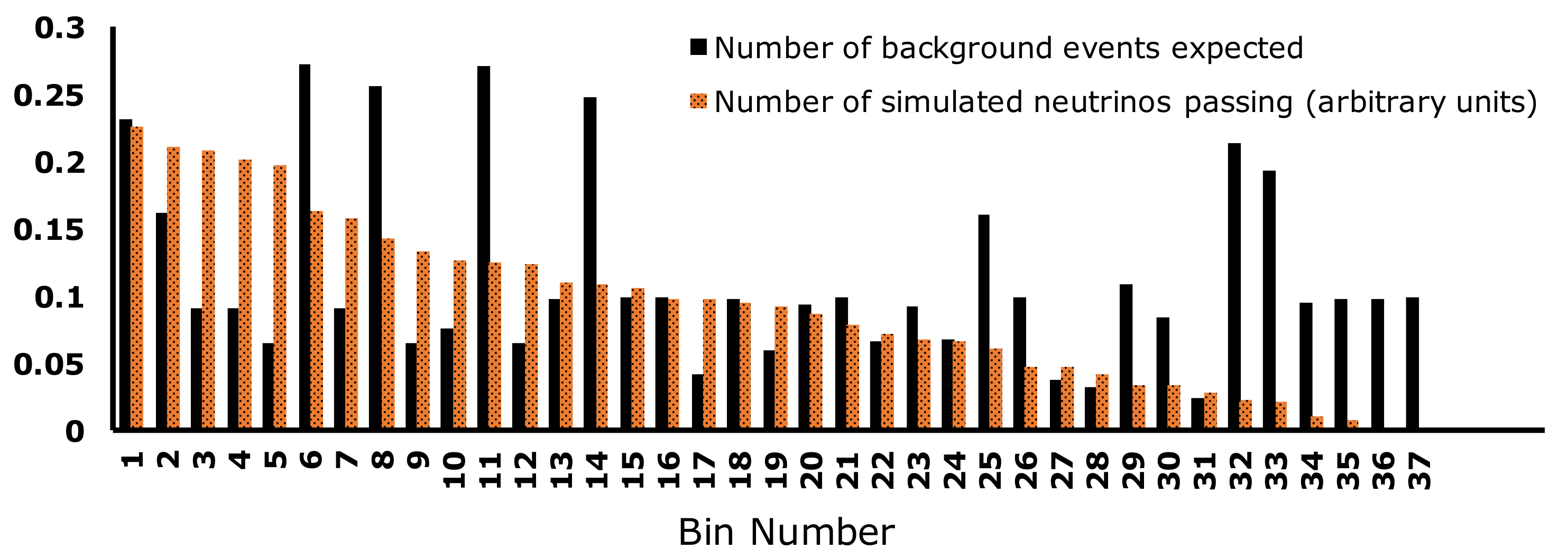}
\caption{ The background estimate and number of expected events from the Kotera model
in each bin of ice used in Analysis{\bf~C}.}
\label{fig:bg_sig_perbin}
\end{figure*}

\subsection{\bf{Ice selection and clustering}} 
\label{sec:binned_clustering}
With the aim of keeping as much ice as possible, in  Analysis {\bf C},
$37$ and $29$ HEALPix bins are kept for the horizontally- and
vertically-polarized (HPol and VPol) channels, respectively, corresponding to
bins that include 61\% and 73\% of the neutrino sensitivity for the VPol and HPol channels, respectively, at the stage where the LD cut is imposed.

In each bin where Analysis {\bf C} performs its search, events from a
10\% sample are used to obtain the background estimate for that bin by fitting to an exponential function in a variable that represents the position of the LD
cut as shown in Fig.~\ref{fig:diffdist}.
Fig.~\ref{fig:bg_sig_perbin} shows the background and signal expectation in 
each bin used.
Bins are removed if there
is not sufficient data beyond the peak to fit (at least five events across five bins).
We also require that the exponential fit returns a p-value between 0.05 and 0.999.  Moreover, we 
rank the bins in order of simulated sensitivity
to neutrinos and place bins in the lowest one
percentile in a sideband region. 

As a final step in the analysis, we reject events that pass all other cuts but cluster with any other horizontally or vertically-polarized events.
In the vertically-polarized
channel, in seven bins, we rejected events that cluster with at least one other event, consistent with
sampling from just 10\% of the data to predict the backgrounds.
Based on the observed clusters in the 
90\% signal region, 
we determined post-unblinding that we expect a 2.2 ``singlet'' (single events not paired with any others during clustering)  background among all VPol and HPol bins, in addition to the pre-unblinding background estimates. 

Table~\ref{tab:osu_effs} shows the efficiencies of the cuts used in
Analysis~{\bf C} on simulated neutrinos
as well as the 90\% dataset.  We note that our quality cuts are less 
efficient compared to the other two analyses (62\%) due to aggressively cutting away regions where satellites are present and in directions where our trigger was masked.  In addition, the analysis cuts for Analysis~{\bf C} rely on high cross-correlation values between waveforms with common fields of view rather than impulsivity, and while this adds a complementarity between analyses, the requirement is on average only 20\% efficient.  This cut varies between bins and is more efficient in some areas.

\subsection{Optimization and systematic uncertainties}
\label{optimization}

We optimize in the VPol and HPol channels separately, in each case for the best constraint
on the chosen neutrino flux, despite only
considering the VPol channel for our Askaryan neutrino search.  
The optimization consists of a few steps, and at every stage, to calculate the expected limit for a given set of LD cuts and orientation, we follow a Bayesian technique with a flat prior on $C$,
a scale factor on the Kotera model, with proper smearing of backgrounds to account for systematic uncertainties. 

\begin{table}[]
\centering
\begin{tabular}{l|l|l|l} 
Total   & 100 & & $5.9\cdot 10^{7}$ \\ \hline
Cut &MC $\nu$ eff.  &Comments& Events  \\ 
 &  cut by each (\%)&  & kept  \\ \hline
Quality cuts  & 62 &  & $3.8\cdot 10^5$ \\
Event selection   &99 & & $1.7\cdot 10^5$  \\
Bin selection   &61 &   & $8.8\cdot 10^{4}$  \\
Final selection  & 19  & range 4.4-29\% &  67   \\ 
Clustering & 98& 7 bins affected& 2 \\ \hline
\end{tabular}
\caption{\label{tab:osu_effs}This table shows neutrino efficiencies
and events kept for each stage of cuts (not cumulative) from the 90\% data set  at each stage of Analysis {\bf C}. Efficiencies are the geometric average from two MC samples, one containing the local noise environment in analysis
chain only, and
another with simulated thermal noise. }
\end{table}

First, 
for a given orientation of the HEALPix map, 
the LD cut for each bin is optimized for the best
constraint on the Kotera model
from that bin alone.  Next, given those
LD cuts, the bins are combined
to obtain an expected limit on $C$, a scale factor for the Kotera model.  Then,
all LD cuts are moved up and down together
for the best expected constraint $C$.
Finally,
all of this is repeated for orientations covering a 10x10 grid in latitude and longitude shifts within one bin increment of
the HEALPix map in order to obtain
the set of LD cuts and orientations
that give the best expected constraint $C$.

For a single bin, in the absence of
systematic uncertainties,
for an expected background 
$b$ and signal events $s$, a 90\% CL upper limit can be placed on the number of signal events $s_{\rm up}$ that satisfies:
\begin{equation} \label{eq:L1}
0.1 < \dfrac{\int_{s_{\rm{up}}}^{\infty} e^{-(b+s)}  (b+s)^b ds}{\int_{0}^{\infty} e^{-(b+s) }  (b+s)^b ds}.
\end{equation}

We can create a combined limit over many bins; the relative number of signal events across bins is set by simulation, and we find a scale factor on the overall flux, $C$, as:
\begin{equation} \label{eq:L}
0.1 <{\dfrac{\int_{s_{\rm{up}}}^{\infty} \prod_{i} \int P_i(b) ~db~ e^{-(b+Cs_i)}  (b+Cs_i)^b ds} {\int_{0}^{\infty} \prod_i \int P_i(b)~db ~e^{-(b+Cs_i) }  (b+Cs_i)^b ds}}.
\end{equation}

Analysis {\bf C} used three different
systematic uncertainties as a function of
cut values
in its optimization procedure \cite{jacobGordonThesis}.  Although these uncertainties undoubtedly
led to conservative cuts with
reduced backgrounds, 
post unblinding we found that these systematic uncertainties were overestimated.  Therefore, we set our measured limit based on statistical errors and uncertainties on exponential fits only. 

The three different uncertainties included
in the optimization
were:  1) errors
on parameters of exponential fit
for modeling the background (calculated to be a few percent), 2)
choice of 
functional form for modeling
background (our dominant uncertainty at tens of percent), and 3) spillover of events
between neighboring 
bins due to imperfect resolution on 
reconstruction direction (negligible for nearly all bins).
To assess the uncertainty due to choice of fit, we 
fit the data in the 10\% sample to
a power law function in each bin, in
addition to the exponential.  

When finding the measured limit on a specific model,
  we account for our estimated singlet background (see Section~\ref{sec:binned_clustering}).
   by generating
  singlets from a Poisson distribution
 with a mean of 2.2$\pm$0.6, and 
 distribute them with equal probability among all bins in the vertical and horizontal polarization channels, consistent with how observed clusters are distributed in Analysis {\bf C}.  
 We set a constraint $C$ that is a factor of 8.7  higher than Analysis {\bf B}.
 However, demonstrating the potential
 for complementing the other analyses,
 25\% of neutrinos seen by Analysis {\bf C} would have been rejected by the others.
 
\subsection{Discussion}
With the aim of maintaining sensitivity 
to neutrinos when viewing ice
where anthropogenic noise is present, Analysis {\bf C} took an
aggressive approach at background mitigation, and together with a choice of complementary analysis variables, has a lower efficiency for the search for a diffuse neutrino flux than the other analyses at present.  For the analysis of data from ANITA-IV, the  focus will be to improve efficiencies without allowing more background to pass.

We note that this approach of searching in bins of ice is well suited for neutrinos from astrophysical transient sources such as gamma-ray bursts. For a given transient, the neutrinos would be expected only from a localized region of ice for a short period of time.
Restricting the search region in both time and 
direction in this way allows for lower analysis thresholds for the same background level.

\newcommand{\effarea}{\ensuremath{\langle A\Omega\rangle(E)}}
\newcommand{\effvol}{\ensuremath{\langle V\Omega\rangle(E)}}

\section{Limit calculation}\label{sec:limit}

The ANITA-III 90\% confidence level limit on the all-flavor-sum diffuse neutrino flux is set by using:
\begin{equation}
\left( \dfrac{Ed^4N}{dE dA d\Omega dt}\right)_{lim} =
\dfrac{s_{up}}{ T \cdot \epsilon_{ana} (E_\nu) \cdot \effarea \cdot \Delta} \, ,
\end{equation}
\noindent
where
$s_{up}$ is the upper side of the Feldman-Cousins confidence interval integrated over the systematic error on the background, for the mean of a Poisson
variable given the number of observed events and the estimated background. The lower side of the Feldman-Cousins confidence interval in our case is 0.  For Analysis B, with one event observed and a background estimate of $0.7^{+0.5}_{-0.3}$, we calculate $s_{up}=3.47$.

The livetime for ANITA-III is $T=17.4$ days, which includes the digitization dead time but not channel masking effects, which are included in the calculation of the acceptance.
$\epsilon_{ana}(E_\nu)$ is the Analysis~\textbf{B} neutrino analysis efficiency as a function of energy.
The experiment acceptance, \effarea, is calculated with a weighted Monte Carlo~\cite{icemc}, and using the approximation:
\begin{equation}
\effarea = \frac{\effvol}{\ell_{{int}}(E)} , 
\end{equation}
where
\effvol is the volumetric acceptance, and $\ell_{{int}}(E)$ is the neutrino interaction length calculated from~\cite{crosssection}.
The values of the ANITA-III acceptance are found in Fig.~\ref{fig:limit}.
Finally, the factor $\Delta=4$ follows the normalization convention in Appendix B of~\cite{rice}, which was derived by considering a typical range in log-energy space of the spectrum that follows a 1/$E$ distribution.

The combined limit for ANITA I-III is calculated using the total
 number of events seen, total expected background, and the
 analysis-efficiency-weighted sum of previously-published effective
 volumes~\cite{anita2}.

\end{document}